\documentclass[prodtf]{nws}
\usepackage{color}
\usepackage[normalem]{ulem}           
\usepackage{url}
\usepackage{natbib}

\usepackage{hyperref}
  \hypersetup{%
    colorlinks,                    
    urlcolor=black,
    linkcolor=black,
    citecolor=black,
    pdfborder={0 0 0},
}

\renewcommand{\journal}[1]{}
\def\PREPRINT{Preprint \pagerange}
\makeatletter
\let\@j@urnal\PREPRINT
\makeatother
\DefineNamedColor{named}{orange}{rgb}{1,0.5,0}
\DefineNamedColor{named}{suppcolor}{rgb}{0.2,0.2,1.0}
\newtheorem{definition}{Definition}

\newcommand{\sbub}[2]{\langle {#1},\,{#2}\rangle}  
\newcommand{\card}[1]{\left|{#1}\right|}        
\newcommand{\kraxn}{\,\,\hbox{\texttt{(\hspace{-0.75em})}}\,\,\,}
\newcommand{\SUPP}[1]{#1}

\title{Superbubbles as an Empirical Characteristic of Directed
  Networks}

\ifx\DOUBLEBLIND\undefined
\author[F.\ G{\"a}rtner et al.]{Fabian G{\"a}rtner\\
  Competence Center for Scalable Data Services and Solutions
  Dresden/Leipzig (scaDS) and Bioinformatics Group, Department of Computer
  Science,
  Universit{\"a}t Leipzig, Augustusplatz 12, D-04107 Leipzig, Germany;
  \\
  Felix K{\"u}hnl\\
  Bioinformatics Group, Department of Computer Science, and
  Interdisciplinary Center for Bioinformatics, Universit{\"a}t Leipzig,
  H{\"a}rtelstra{\ss}e 16--18, D-04107 Leipzig, Germany;
  \\
  Carsten R.\ Seemann\\
  Bioinformatics Group, Department of Computer Science, and
  Interdisciplinary Center for Bioinformatics, Universit{\"a}t Leipzig,
  H{\"a}rtelstra{\ss}e 16--18, D-04107 Leipzig, Germany;
  Max Planck Institute for Mathematics in the Sciences, Inselstra{\ss}e 22,
  D-04103 Leipzig, Germany; \email{carsten@bioinf.uni-leipzig.de}
  \\
  The Students of the Graphs and Networks Computer Lab 2018/19\\
  Bioinformatics Group, Department of Computer Science Universit{\"a}t
  Leipzig, H{\"a}rtelstra{\ss}e 16--18, D-04107 Leipzig, Germany;
  \\
  Christian H{\"o}ner zu Siederdissen\\
  Bioinformatics Group, Department of Computer Science, and
  Interdisciplinary Center for Bioinformatics, Universit{\"a}t Leipzig,
  H{\"a}rtelstra{\ss}e 16--18, D-04107 Leipzig, Germany;
  \\
  Peter F.\ Stadler\\
  Bioinformatics Group, Department of Computer Science; Interdisciplinary
  Center for Bioinformatics; Competence Center for Scalable Data Services
  and Solutions Dresden/Leipzig (scaDS); German Centre for Integrative
  Biodiversity Research (iDiv) Halle-Jena-Leipzig; and Leipzig Research
  Center for Civilization Diseases, Universit{\"a}t Leipzig,
  H{\"a}rtelstra{\ss}e 16--18, D-04107 Leipzig, Germany; Max Planck
  Institute for Mathematics in the Sciences, Inselstra{\ss}e 22, D-04103
  Leipzig, Germany; Institute for Theoretical Chemistry, University of
  Vienna, W{\"a}hringerstra{\ss}e 17, A-1090 Wien, Austria; Facultad de
  Ciencias, Universidad National de Colombia, Sede Bogot{\'a}, Colombia;
  Santa Fe Institute, 1399 Hyde Park Rd., Santa Fe, NM 87501, USA;
  \email{\{fabian,felix,carsten,choener,studla\}@bioinf.uni-leipzig.de}
  }
\else
\author[]{}
\fi

\jdate{Febrember 2099}
\pubyear{2099}
\pagerange{\pageref{firstpage}--\pageref{lastpage}}
\doi{S0956796801004857}

\begin{document}
\label{firstpage}
\maketitle

\begin{abstract}
  Superbubbles are acyclic induced subgraphs of a digraph with single
  entrance and exit that naturally arise in the context of genome
  assembly and the analysis of genome alignments in computational
  biology. These structures can be computed in linear time and are
  confined to non-symmetric digraphs. We demonstrate empirically that
  graph parameters derived from superbubbles provide a convenient means
  of distinguishing different classes of real-world graphical models,
  while being largely unrelated to simple, commonly used parameters.
\end{abstract}

\tableofcontents

\section{Introduction}

Directed networks play an important role in real-world graphical
models. Well-documented examples include hyperlinks connecting web pages,
prey--predator relationships, dependencies in project management and
scheduling, or chemical reaction networks. Undirected networks can be seen
as special case of directed ones, with each undirected edge corresponding
to a pair of arcs in opposite direction (symmetric digraphs). Road
networks, for instance are directed but quite close to symmetric as only a
minority of all roads are one-ways. Although there are many properties of
directed graphs that reduce to uninteresting trivialities in the case of
symmetric/undirected graphs \citep{BangJensen:09}, the most commonly used
quantitative characteristics of graph structure, such as density, distance,
betweenness centrality, clustering coefficients, etc.\ \citep{Barabasi:16},
are measures that are applicable in essentially the same way to both
directed and undirected networks.

\begin{figure}[b]
\begin{tabular}{rcl}
\begin{minipage}{0.5\textwidth}
  \noindent\includegraphics[width=\textwidth]{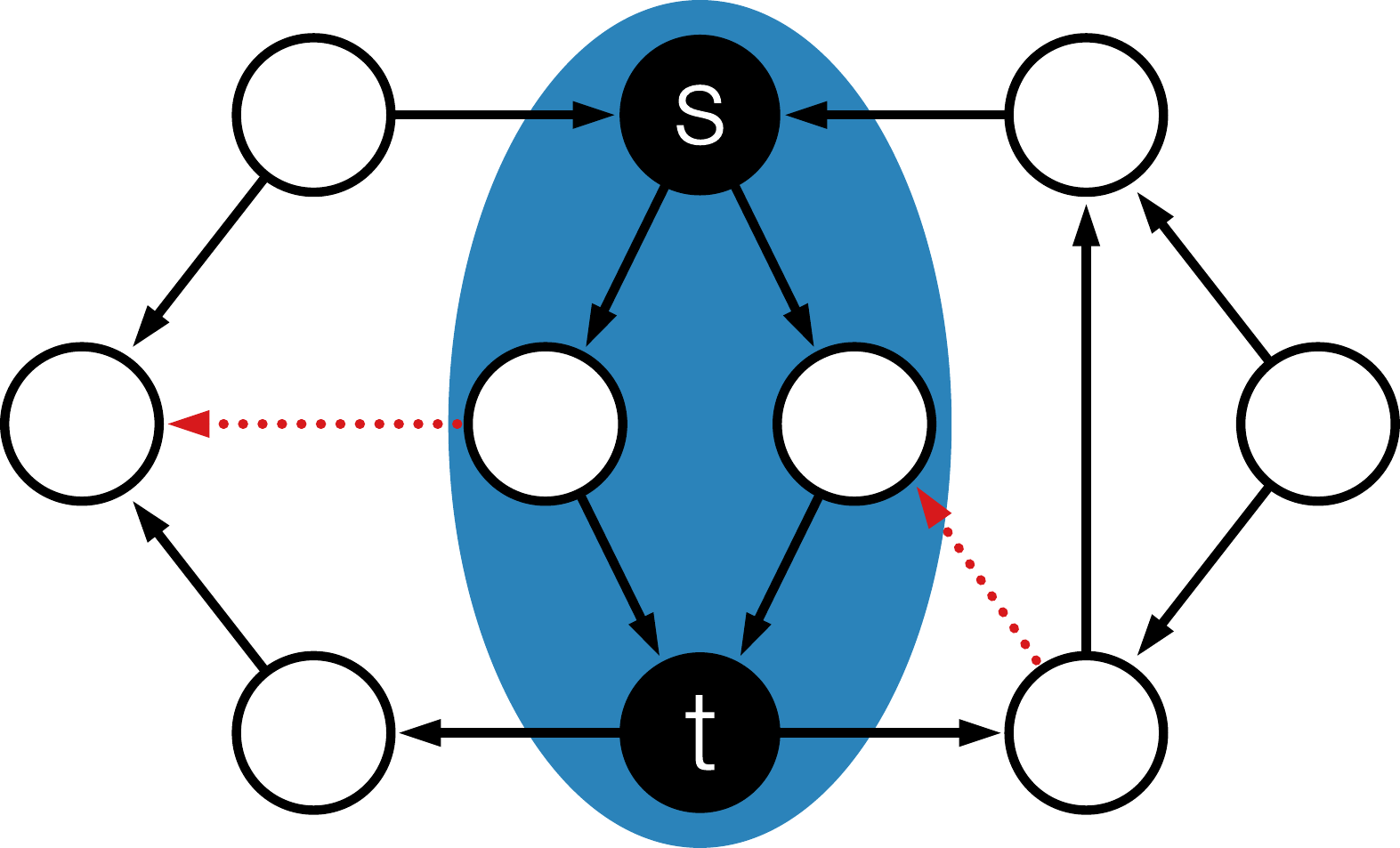}
\end{minipage}
&\hspace*{0.5cm}&
\begin{minipage}{0.45\textwidth}
  \caption[Definition of superbubbles]{%
    A superbubble $\sbub{s}{t}$ (shaded area) is an acyclic induced
    subgraph with a single entrance $s$ and a single exit $t$ such that
    every vertex in $\sbub{s}{t}$ is reachable from $s$ and reaches $t$.
    Directed edges leading into or out of interior vertices of $\sbub{s}{t}$
    are forbidden (red dotted edges). In addition, a superbubble is a
    minimal induced subgraph with given entrance $s$ or exit $t$.
  }
  \label{fig:def-sb}
\end{minipage}
\end{tabular}
\end{figure}

Therefore, it is interesting to ask whether there are quantitative
characteristics that are in a sense particular to digraphs. Conceptually,
these capture properties of digraphs that have no (interesting) analog in
undirected networks. Of course, there are some simple measures of this
type, such as the degree imbalance $d^+(v)-d^-(v)$ \citep{Mubayi:01} or
directional difference $d(x,y)-d(y,x)$ of the length of shortest paths. All
the examples that we are aware of, however, are ``very local'' in their
nature.

Bubble structures in digraphs recently have attracted interest in
computational biology, where they identify module-like features in genome
assembly graphs that can be processed independently \citep{Paten:18}.
\begin{definition}[Superbubble; \citet{Onodera:13,Sung:15}]
  A \emph{superbubble} $S=\sbub{s}{t}$ in a digraph $G$ is a minimal,
  acyclic, induced sub-digraph with a single \emph{entrance} $s$ and a
  single \emph{exit} $t$ such that (i) no vertex in $S$ can be reached from
  the outside without passing through $s$, (ii) no vertex outside from $S$
  is reachable from within $S$ without passing through $t$, (iii) every
  vertex within $S$ is reachable from $s$ and can reach $t$.  The vertices
  in $V(\sbub{s}{t})\setminus\{s,\,t\}$ are referred to as the
  \emph{interior} vertices of $\sbub{s}{t}$.
\end{definition}
In particular, for a superbubble $\sbub{s}{t}$ there are no induced
subgraphs of the form $\sbub{s}{t'}$ or $\sbub{s'}{t}$, with $t'\ne t$ or
$s'\ne s$, resp., that satisfy conditions (i), (ii), and (iii). The
definition is illustrated in Fig.~\ref{fig:def-sb}. It follows directly
from the definition that two superbubbles can only intersect if either one
is properly contained in the other or the exit of one serves as the
entrance of the other. As a consequence, the number of superbubbles in a
digraph cannot exceed the number of vertices.

By definition symmetric digraphs do not contain induced acyclic connected
subgraphs with two or more vertices and thus do not contain
superbubbles. Since all superbubbles in a digraph $G$ can be identified and
listed in linear time \citep{Brankovic:16,Gaertner:18b,Gaertner:19a}, they
potentially serve as a genuine characteristic of directed features in $G$.
The purpose of this contribution is to demonstrate empirically that
superbubbles
indeed provide useful quantitative parameters to describe digraphs.

\section{Methods}

\subsection{Enumeration of all Superbubbles}

\begin{figure}
  \begin{tabular}{rcl}
    \begin{minipage}{0.58\textwidth}
    \includegraphics[width=\textwidth]{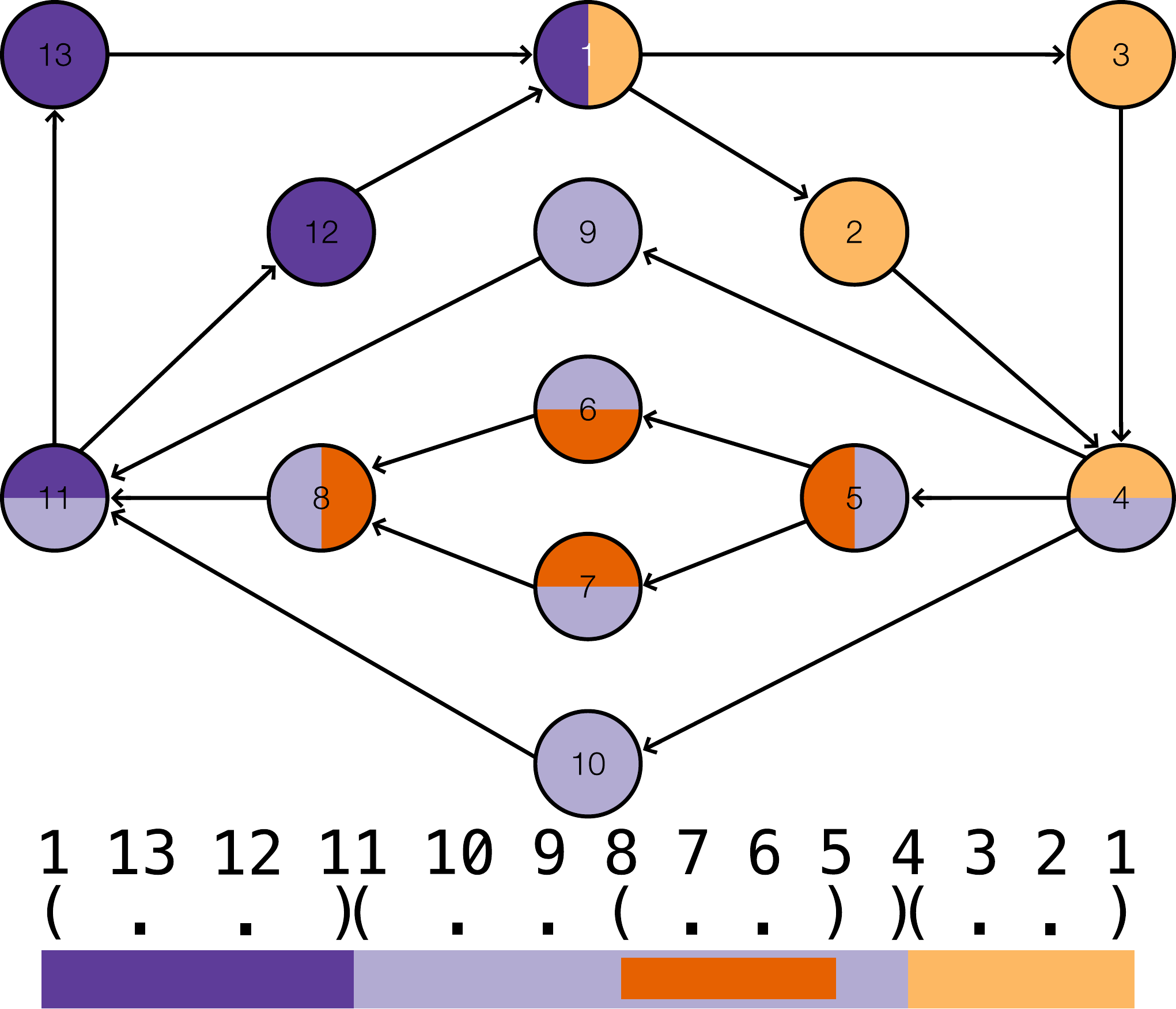}
  \end{minipage}
    &\hspace*{0.04\textwidth}&
  \begin{minipage}{0.35\textwidth}
    \caption[Nested supberbubbles]{A graph $G$ with four superbubbles
      (indicated by the coloring of the vertices). Below, superbubbles are
      annotated as matching pairs of parentheses in the DFS order of the
      vertices used by \texttt{CLSD} to compute them. The symbols
      \texttt{(} and \texttt{)} denote the exit and entrance,
      respectively. In this example, every vertex of $G$ is an exit or an
      interior vertex, hence the root $1$ of the DFS search appears twice
      (see text). The vertices \texttt{11} and \texttt{4}, marked by symbol
      \kraxn, are the entrance of one and the exit of another superbubble.}
    \label{fig:sbstring}
  \end{minipage}
  \end{tabular}
\end{figure}

A key feature of superbubbles is that their vertices appear as an interval
in the postorder of certain DFS trees
\citep{Brankovic:16,Gaertner:18b,Gaertner:19a}. Although they may be
nested, two superbubbles cannot have the same entrance or the same exit
vertices \citep{Onodera:13}.  Hence, the number of superbubbles lies in
$O(|V|)$.  Furthermore, it is easy to read off the nesting pattern of
superbubbles from the DFS-post-ordered vertices, when entrance and exit
vertices are marked by matching pairs of parentheses. Since the entrance of
a superbubble can be the exit of another one, the extra symbol \kraxn is
used for vertices that are both exit and entrance, as shown in
Fig.~\ref{fig:sbstring}.

To detect superbubbles, we use the algorithm of \cite{Gaertner:19a}. First, it
identifies a set of roots for DFSs. Then, every superbubble appears as
an uninterrupted interval in the postorder of the forest composed of the
DFS trees. This makes it easy to compute the size of the superbubbles as
well as their nesting patterns in linear time. To list all superbubbles,
the initial vertex of the DFS search must not be an exit or interior vertex
of a superbubble. There are, however, graphs such as the example in
Fig.~\ref{fig:sbstring} for which every vertex is an exit or interior point
of a superbubble. In such cases an auxiliary graph is analyzed in which an
exit $t$ is split into two, with one copy only retaining the incoming and
the other retaining only the outgoing edges. We refer to
\cite{Gaertner:19a} for a detailed description of the algorithm.

In order to provide an efficient tool to determine the superbubbles in
large digraphs, we reimplemented the \texttt{python} toolkit
\texttt{LSD} in \texttt{C++}. The \texttt{CLSD} software not only
substantially improves performance, it also provides modules to compute
various summary statistics. It is available at
\texttt{github}.\footnote{\url{https://github.com/Fabianexe/clsd}}

\subsection{Superbubble Descriptors}

Superbubbles form the basis of a rich set of graph descriptors. The
simplest quantities are the number of superbubbles, their total size, and
measures such as the number of vertices or edges contained in a
superbubble. These quantities are naturally normalized by the number of
vertices.

Since superbubbles are induced acyclic subgraphs, a wide variety of
conventional graph descriptors can be computed from them. The density of a
superbubble $S$, for example, is defined by
$\rho(S):=\frac{2\card{E(S)}}{\card{V(S)}(\card{V(S)}-1)}$, where $E(S)$
and $V(S)$ is the edge set and vertex set of $S$, respectively.

Recalling that superbubbles are computed with the help of a special DFS
tree $T$. Note that several interesting quantities can be computed very
efficiently. The reverse finishing order of a DFS equals the postorder of a
DFS tree $T$ \citep{Gaertner:19a}. In an acyclic graph, the postorder
implies a topological sorting of the graph \citep{Tarjan:72}. Therefore, it
is possible to compute the length $\ell$ of the longest path as well as the
number $n$ of the distinct paths within each superbubble in linear time. To
this end, we start at the exit $u$ of a superbubble with $n_u=\ell_u=0$ and
propagate the number of paths in postorder to the entrance:
\begin{equation*}
  n_x = \sum_{y\in N^<(x)} n_y \qquad\textnormal{and}\qquad
  \ell_x = \max_{y\in N^<(x)} \ell_y
\end{equation*}
where $N^<(x)$ is the set of neighbors that precede $x$ in postorder.
One easily checks that $\ell_x\le \card{V(S)}$ and
$n_x\le 2^{\card{V(S)}-2}$.

\begin{figure}
  \begin{center}
    \includegraphics[width=0.8\textwidth]{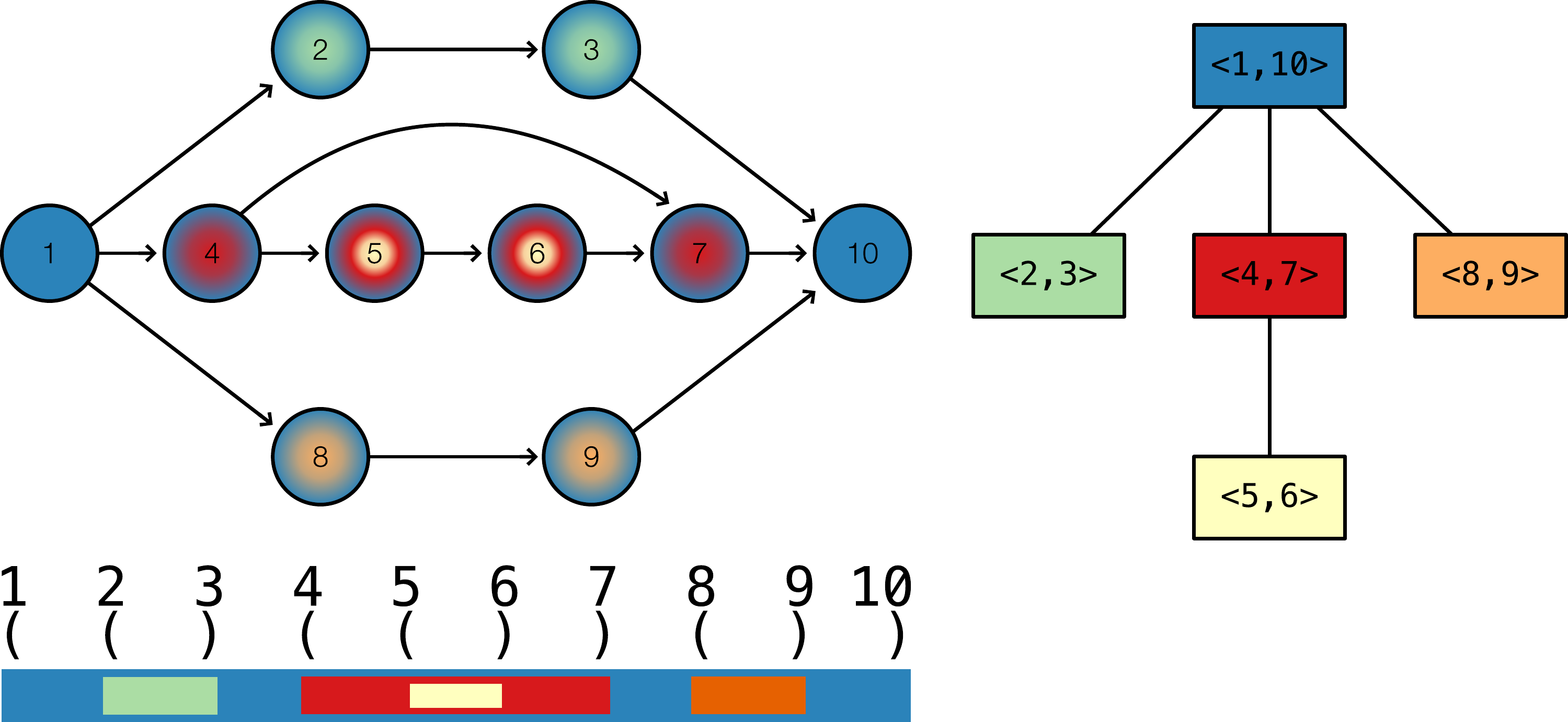}
  \end{center}
  \caption[Example of a superbubble hierarchy]{Example of a superbubble
    hierarchy and the tree that represents its nesting structure. To the
    left, there is a graph with one superbubble hierarchy with five
    superbubbles. Below, a DFS postorder of the hierarchy where the
    superbubbles are marked with parentheses and colored bars is shown. To
    the right, the superbubble tree that corresponds to the hierarchy is
    depicted. It is rooted on the superbubble $\sbub{1}{10}$. Thus, the
    hierarchy has a maximal depth of three and contains five superbubbles.
  }
  \label{fig:nestingtree}
\end{figure}

\begin{definition}[(Non-)trivial Superbubble]\label{def:trivial-sb}
  A superbubble consisting only of the entry and the exit node is called
  \emph{trivial}, and \emph{non-trivial}, otherwise.
\end{definition}
We consider trivial superbubbles as a special case
since they usually appear much more frequently in many networks of
different types. We therefore do no include them in the computation of
average properties of superbubbles.

\begin{definition}[Superbubble Hierarchy]
  A \emph{superbubble hierarchy} is a set $\mathcal{S}$ of superbubbles
  comprising exactly one inclusion-maximal superbubble $S$ and all
  superbubbles $S'\subset S$.\\
  We say $\mathcal{S}$ is \emph{flat} if it consists of a single
  superbubble and \emph{nested} if $|\mathcal{S}|\ge 2$.
\end{definition}
Recall that for two distinct superbubbles $S_1$ and $S_2$ exactly one of
the following three alternatives is true: $S_1\subset S_2$,
$S_2\subset S_1$, or $|S_1\cap S_2|\le 1$. In the last case, no superbubble
can be contained in $S_1\cap S_2$. Therefore, every superbubble is
contained in exactly one \emph{superbubble hierarchy}. Although a
superbubble hierarchy $\mathcal{S}$ in general does not satisfy the usual
axioms for hierarchical set systems, it is still true that the Hasse
diagram of $\mathcal{S}$ w.\,r.\,t.\ to set inclusion is a tree. The
corresponding nesting of the superbubbles is easy to identify in the DFS
postorder, see Fig.~\ref{fig:sbstring}: $\sbub{s'}{t'}$ is included in
$\sbub{s}{t}$ if the parentheses representing $\sbub{s}{t}$ enclose those
of $\sbub{s'}{t'}$. The Hasse diagram is faithfully captured by
re-interpreting the pairs of parentheses as nodes in a forest, see
Fig.~\ref{fig:nestingtree}.  The roots of the constituent trees of the
forest are the inclusion-maximal superbubbles, while inclusion-minimal
superbubbles appear as leaves.

We use the number of superbubbles in each hierarchy $\mathcal{S}$,
i.\,e., each tree, as well as the depth of the trees as convenient
descriptors. The trees are easily extracted from the DFS-based string
representation in linear time in a single pass: whenever an exit is
encountered, a superbubble is pushed onto the stack and recorded as child
of the superbubble previously on top of the stack (or a root, if the stack
was empty). When an entrance is found, the superbubble is popped from the
top of the stack. A constituent tree is completed whenever the stack is
empty.

In the following, we use 14 quantities based on superbubbles: the number of
superbubbles (\texttt{S}), the fraction of vertices and edges that are in a
superbubble (\texttt{VS} and \texttt{ES}), the number of trivial
superbubbles (``mini'', \texttt{MS}), the maximum number of vertices and
edges that a single superbubble has (\texttt{mVS} and \texttt{mES}), the
number of superbubble hierarchies (``complexes'', \texttt{C}), the largest
number of superbubbles in one hierarchy (\texttt{CS}), the maximum depth
that a single superbubble has (\texttt{depth}), the maximum number of paths
and path length in one superbubble (\texttt{P} and \texttt{PL}), and the
average values of number of paths, path length, and density (\texttt{aP},
\texttt{aPL}, and \texttt{SD}) for non-minimal superbubbles.

\subsection{Other Graph Descriptors}

The published literature discusses a plethora of graph descriptors, many of
which were designed to parametrize ``quantitative structure--activity
relationships'' (QSAR) in chemistry \citep{Devillers:00}. The overwhelming
majority, however, describes features of undirected graphs. In contrast to
applications in chemistry, where molecular graphs are typically small,
comprising maybe a few hundred nodes, we are interested here in very large
directed networks. We therefore consider only descriptors that can be
computed in (nearly) linear time. In particular, this rules out the
different centrality measures
\citep{Sabidussi_1966,Hage_1995,Shimbel_1953,Brandes_2001,Anderson_1985}.

In order to investigate the scaling of graph descriptors, we of course
record the basic measures of graph size, i.\,e., the number of vertices
$N$, number of edges $M$, and the graph density
$\texttt{GD}=\frac{M}{N(N-1)}$.  Other basic descriptors are the number of
connected components \texttt{CC} and the maximal vertex degree $\Delta$. Two
commonly used measures derived from the degrees are the
assortativity~\texttt{R}~\citep{Newman_2003}, which measures the correlation
of the degrees of adjacent vertices, and the normalized
self-similarity~\texttt{SS}~\citep{Li_2005}.

Vertex degrees are naturally divided into in- and out-degrees for digraphs.
Hence, we consider the largest in-degree ($\Delta_\leftarrow$), the largest
out-degree ($\Delta_\rightarrow$), and the number of non singular strongly
connected components (\texttt{SCC}), and the fraction of bi-directional
edges. If the latter approaches $1$, then the digraph becomes equivalent to an
undirected graph, and no superbubbles can be present. A directed version of
the assortativity was introduced in \cite{Foster_2010} considering the
correlation between in- and out-degrees of adjacent vertices, which appears
in four combinations $R_{\leftarrow\leftarrow}$,
$R_{\leftarrow\rightarrow}$, $R_{\rightarrow\leftarrow}$, and
$R_{\rightarrow\rightarrow}$.  Finally, we consider the heterogeneity
index~$\texttt{H}$~\citep{Ye_2013}, which can be considered as a variant of
the directed assortativities.

\subsection{Datasets}

We used three standard random network models to generate the first
dataset. In the Erd{\H{o}}s--R{\'e}nyi model, directed edges are inserted
independently with a probability $p$ that corresponds to the expected graph
density \citep{Erdoes:59}. The Barab{\'a}si--Albert model
\citep{Barabasi:99,Bollobas:03} uses preferential attachment depending in a
natural way on the in- and out-degrees.  In the Watts--Strogatz model, a
regular ring lattice is rewired with a given per-edge probability, see
\cite{Watts:98,Bollobas:03}.

Here, we compare directed graphs with $N=10\;000$ vertices each. For the
Erd{\H{o}}s--R{\'e}nyi model, an insertion probability of $p = 0.05$ was
used. In the Barab{\'a}si--Albert model, the attachment probability was chosen
to scale linearly with the vertex degree. The Watts--Strogatz model was set to
use a neighborhood of size $2\times5$ during the ring lattice construction,
and a rewiring probability of $0.05$.

\emph{LDBC Graphalytics}\footnote{\url{https://graphalytics.org/datasets}}
offers benchmark data sets primarily intended for the comparison
of graph analysis platforms. These graphs were simulated using different
elaborate random models and were originally
intended as undirected graphs. We obtained directed versions by
re-interpreting the edge lists as directed edges. Three distinct models
were used: the \textit{graph500} data set (\texttt{g500}) \citep{Murphy:10},
the ``Facebook model'' (\texttt{fb}) \citep{Capot__2015}, and a class of
graphs with Zipf-distributed vertex degrees (\texttt{zf})
\citep{METCALF201667}.

The \emph{Stanford Large Network Dataset Collection} \citep{Leskovec:16}
provides a wide variety of empirical network data for
download.\footnote{\url{http://snap.stanford.edu/data/}}  We used here a
subset of the directed networks.

As further real-life examples we investigated the ``supergenome graphs''
constructed from multiple alignments of related genomes. These encode the
rearrangements and other changes of genomes during evolution at larger
scales than insertion, deletion, and substitution of individual nucleotides
(letters), see \cite{Herbig:12,Gaertner:18a}. The graphs used here were all
taken from \cite{Gaertner:18a}.

\section{Results}

\subsection{Random Graph Models}

The simple random models behave very differently. While neither the
directed Erd{\H{o}}s--R{\'e}nyi graphs nor the directed Watts--Strogatz small
world networks contain any superbubbles, they are abundant in the
Barab{\'a}si--Albert preferential attachment graphs. With the parameter
settings outlined in the Methods part we find about $0.3$ superbubbles per
vertex. No non-trivial superbubbles were observed. Among the LDBC models
investigated here, the \texttt{zf} set shows by far the largest density of
superbubbles (about $0.06$ per vertex). These occasionally contain
non-trivial superbubbles. A much smaller number of about $0.003$
superbubbles per vertex was found in the \texttt{g500} set, while the
\texttt{fb} model rarely generates superbubbles at all. Taken together, the
analysis of the random graphs models shows that the density of superbubble
is a sensitive measure that picks up -- sometimes subtle -- differences
between random graph models.

\subsection{Analysis of Real-World Network Data}

Most of the graphs from the \emph{Stanford Large Network Dataset
Collection} contain superbubbles. However, comparably large values, 
above $0.01$ per vertex, seem to be rare. On the other hand, only 4
of 37 graphs were devoid of superbubbles.

\begin{figure}
  \begin{center}
    \includegraphics[width=0.48\textwidth]{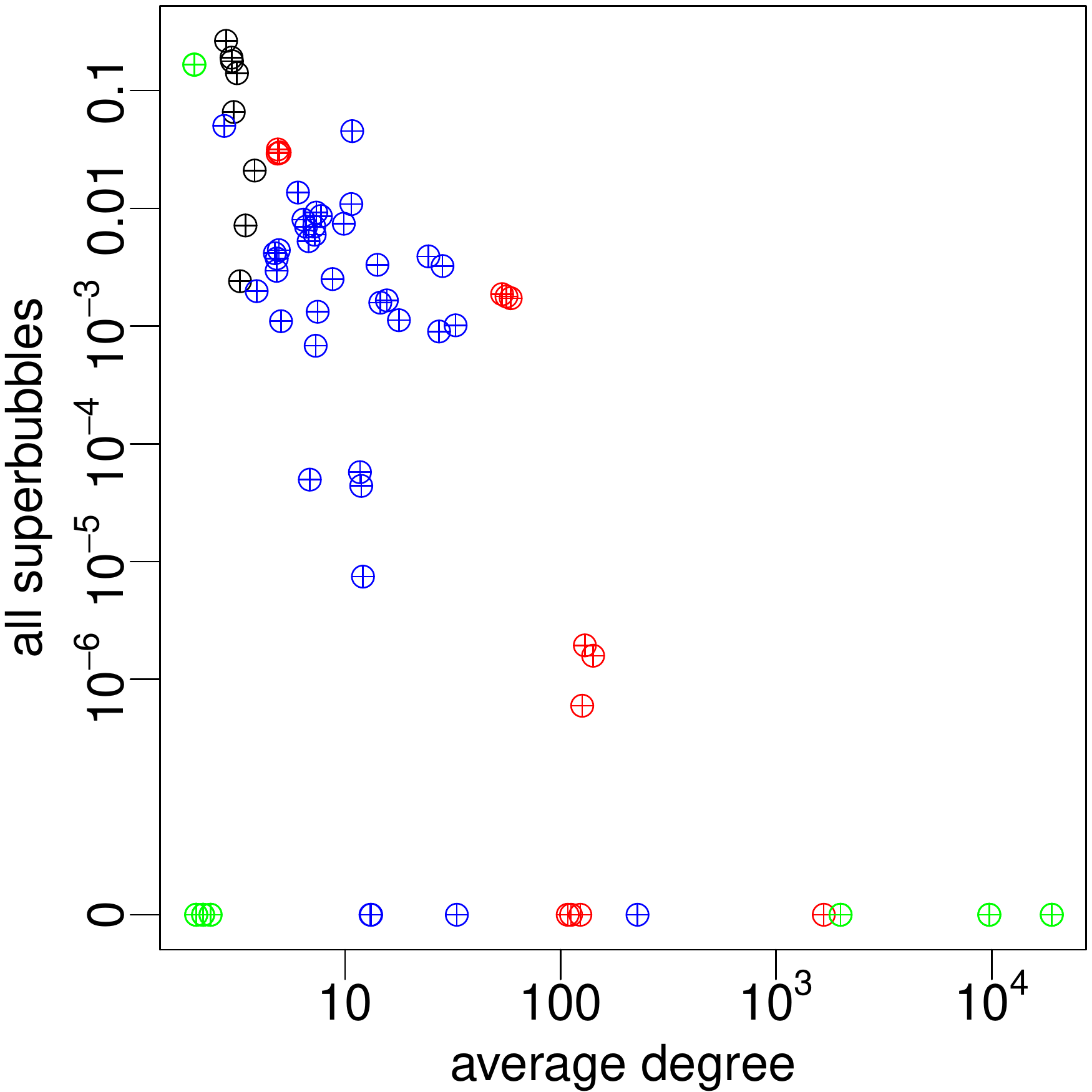}%
    \hspace{0.04\textwidth}%
    \includegraphics[width=0.48\textwidth]{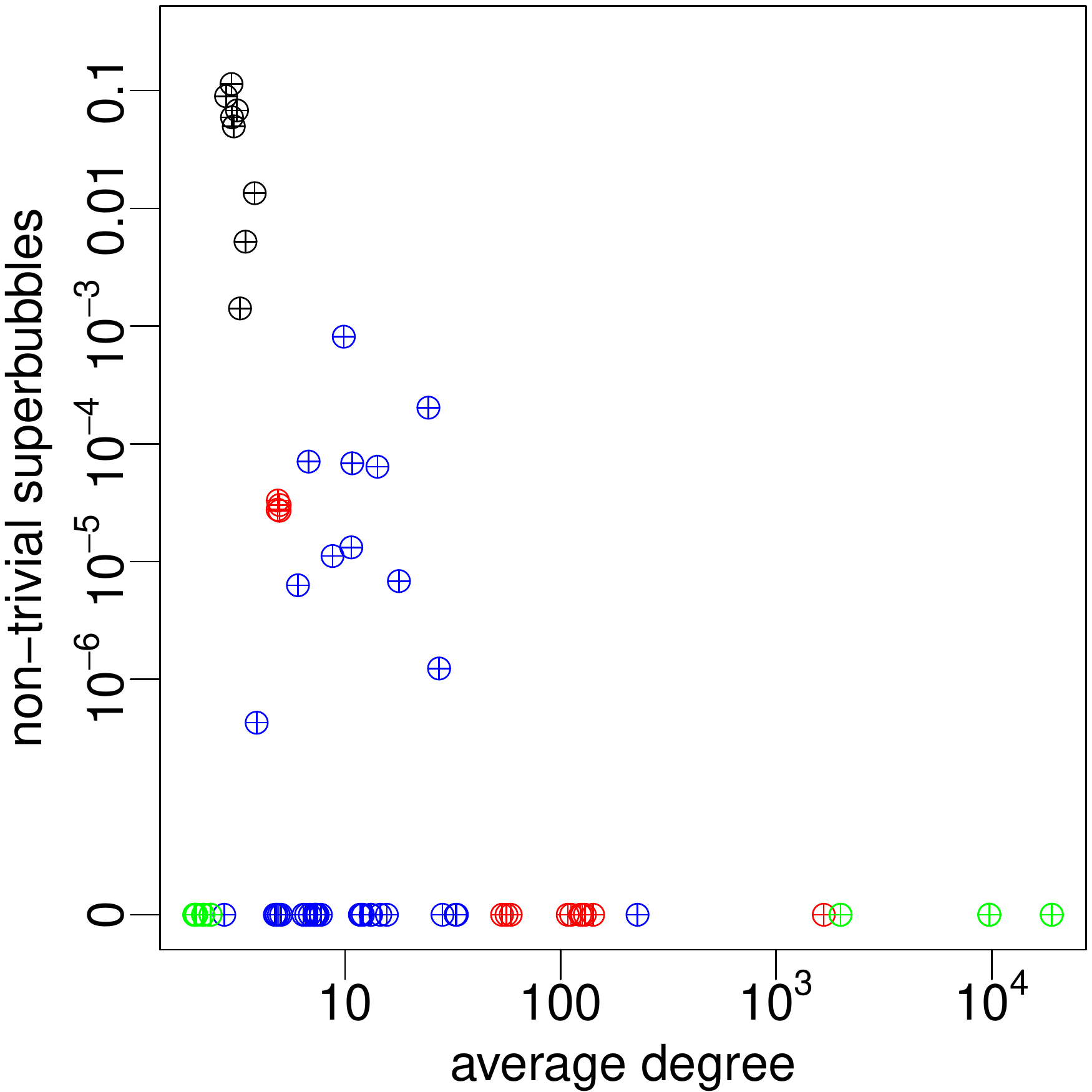}
  \end{center}
  \caption[Supperbubbles in multiple datasets]{%
    Normalized number of superbubbles as a function of and average degree
    in several artificial and real-world network datasets. The left-hand
    side shows \emph{all} superbubbles, the right-hand side only
    \emph{non-trivial} superbubbles. The datasets are colored as follows:
    black: supergenome graphs, red: LDBC dataset, blue: Stanford dataset,
    green: standard random models dataset.}
  \label{fig:bubble_fraction_all}
\end{figure}

The datasets
email-EuAll\footnote{\url{http://snap.stanford.edu/data/email-EuAll.html}}
and
soc-sign-epinions\footnote{\url{http://snap.stanford.edu/data/soc-sign-epinions.html}}
are notable because their fractions of vertices in superbubbles are four to
five times higher than those of the other datasets ($0.100$ and $0.089$,
respectively). However, these are, as for most of the datasets, trivial
superbubbles in the majority of the cases. The largest fraction of vertices
in non-trivial superbubbles is contained in the
web-Google\footnote{\url{http://snap.stanford.edu/data/web-Google.html}}
dataset and is 0.0025. For a comparison of the networks from all datasets,
we refer to Fig.~\ref{fig:bubble_fraction_all}.

There is no clear relation of the frequency of (non-trivial) superbubbles
with the average degree, Fig.~\ref{fig:bubble_fraction_all}, or other
common measures, although denser digraphs naturally tend to have more small
directed cycles, and thus fewer and smaller superbubbles. We also observe
some clustering association of superbubble abundance with the type of graph
generator, with supergenome graphs having the highest incidence of
superbubbles.

Nested superbubble hierarchies seem to be exceedingly rare in both the
usual random network models and social networks. We only occasionally
encountered a hierarchy of depth $2$ in these data, see \SUPP{Supplemental
  Tables 6 and 12}.


% ...

% ...

\subsection{Applications in Sequence Analysis}

Superbubbles were first described as features of \emph{assembly graphs} in
computational biology \citep{Onodera:13}. These graphs describe an
intermediate stage in the reconstruction of genomic sequences from short
experimentally determined sequences. A closely related class of graphs
arises from genome-wide multiple sequence alignments. Here, each vertex
denotes a so-called alignment block, i.\,e., a collection of intervals of
genomic sequences from different species that correspond to each
other. Directed edges keep track of the linear order of sequences within
each of the genomic DNA sequences under consideration
\citep{Herbig:12,Gaertner:18a}.

\begin{figure}
  \begin{tabular}{lcr}
    \begin{minipage}[c]{0.40\textwidth}
      \includegraphics[width=\textwidth]{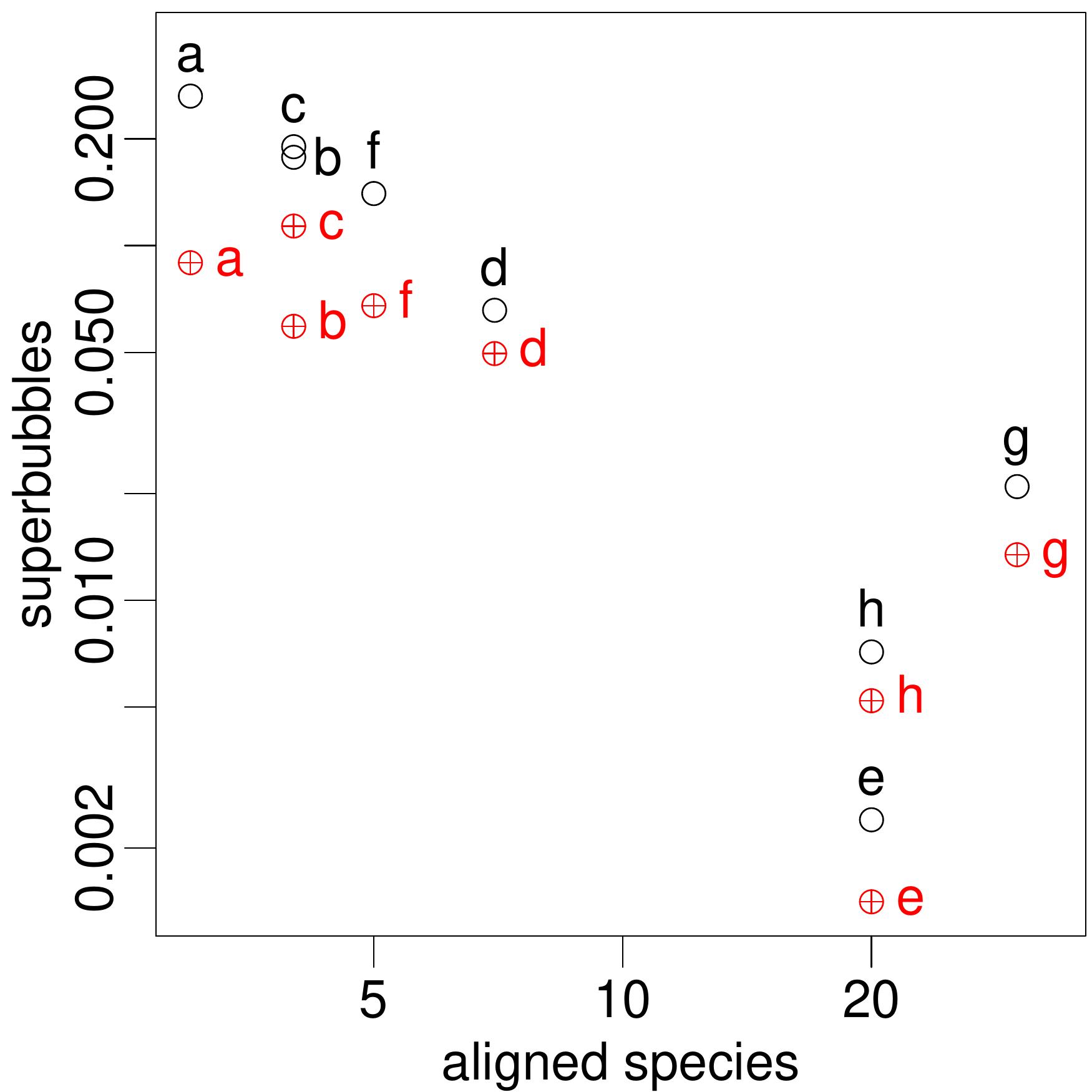}
    \end{minipage}
    & 
    & \begin{minipage}[c]{0.45\textwidth}\small
  \begin{tabular}{cllrlr}
    &\multicolumn{3}{c}{\textbf{Alignment}}
    &\multicolumn{2}{c}{\textbf{Superbubbles}}\\
    a & Dog     & \texttt{canfam1\_3}  & 3  & 0.415 & 0.289 \\
    b & Dog     & \texttt{canfam2\_4}  & 4  & 0.311 & 0.197 \\
    c & Cat     & \texttt{felcat3\_4}  & 4  & 0.406 & 0.345 \\
    d & Finch   & \texttt{geofor1\_7}  & 7  & 0.161 & 0.151 \\
    e & Human   & \texttt{hg38\_20}    & 20 & 0.006 & 0.005 \\
    f & Mouse   & \texttt{mm5\_5}      & 5  & 0.301 & 0.223 \\
    g & Mouse   & \texttt{mm9\_30}     & 30 & 0.050 & 0.039 \\
    h & Tarsier & \texttt{tarsyr2\_20} & 20 & 0.019 & 0.015 \\
  \end{tabular}
  \vspace{2em}    
\end{minipage}
\end{tabular}
\caption[Superbubbles in supergenome graphs]{%
  Normalized number of superbubbles in supergenome graphs. Black circles
  refer to \emph{all} superbubbles, red crossed circles denote only
  \emph{non-trivial} superbubbles. Each supergenome graph is derived from a
  multiple alignment for which we list the the reference species, the
  ENSEMBL designation, and the number of species, as well as the normalized
  number of all and the non-trivial superbubbles. Note the $\log$--$\log$
  scale of the graph.}
\label{fig:supergenome}
\end{figure}

Fig.~\ref{fig:supergenome} shows that these digraphs, which have sizes
between $3\;783\;877$ and $30\;368\;906$ vertices, contain a comparatively
large number of superbubbles. A large fractions (between 33\% and 75\%) of
the superbubbles in each graph are of the non-trivial type. In
comparison, only a few of the social networks discussed above contain
2--11\% of non-trivial superbubbles, with less than 1\% in most of the
examples we analyzed. In supergenome graphs, usually more vertices are
covered by non-trivial superbubbles than by trivial ones (the only
exception are the two Dog-centered alignments). There is an overall
negative correlation of the superbubble abundance with the number of species
included in the alignments. An increase in the number of species implies
an increased number of genome rearrangements detectable in the alignment,
which in turn increases the abundance of cycles. Intuitively, it is
plausible that acyclic induced subgraphs -- and thus also 
superbubbles -- become less abundant.

Nested superbubble hierarchies are also more abundant in the supergenome
graphs than in any other class of digraphs we have investigated. While
still rare compared to flat superbubbles, we find examples with a depth up
to $6$ in graphs derived from alignments of both few and many species, see
\SUPP{Supplemental Table 3.}

\section{Concluding Remarks}

Superbubbles were introduced here as a means to simplify graphs arising in
genome assembly and related applications in computational
biology. Since superbubbles are connected acyclic induced subgraphs,
one can expect a negative relationship with the abundance of small
cycles. However, superbubbles and short cycles do not convey the same
information. In particular, superbubbles may also distinguish classes of
acyclic graphs, which may have anywhere between no superbubbles at all
and being completely covered by them. Maybe even more importantly, the
list of all superbubbles in a directed graph can be computed in linear
time, and thus they can be obtained and evaluated for very large networks.
The enumeration of short cycles (with length $\ell\le7$) takes
$O(|V|^{\omega})$ time and $O(|V|^2)$ space \citep{Alon:97}, where
$\omega\approx 2.3729$ is the ``matrix multiplication constant'', which
will in general not be feasible for very large sparse graphs. Only
triangles can be enumerated in $O(|E|^{\frac{2\omega}{\omega+1}})$.

In contrast to most other commonly used descriptors of digraphs,
superbubbles are genuinely a feature of digraphs in the sense that (1) no
associated construction exists for undirected graphs and (2) superbubbles
do not exist in symmetric digraphs. In this contribution, we showed that
large numbers of superbubbles appear in directed networks with
Zipf-distributed vertex degrees as well as directed versions of
preferential attachment-based graphs. They also appear in many real-world
graphs and seem to be sensitive to structural features that are not
captured well by parameters that depends on local vertex degrees. While
superbubbles are abundant in many graph classes, we find that nested
superbubble hierarchies, i.\,e., treelike structures composed of
superbubbles, appear only in a few special graph classes. They are most
abundant in the graphs deriving from genome alignments.

The empirical usefulness of superbubbles as a means of quickly computing
digraph-specific numerical descriptors suggests to investigate the
theoretical distributions of superbubbles -- and possibly related locally
acyclic structures -- in various random graph models. The clear
differences between classes of social network graphs and supergenome
graphs, for examples, also suggests that it will be a worthwhile effort
to develop random digraph models that exhibit large numbers of
non-trivial superbubbles and nested superbubble hierarchies.


\ifx\DOUBLEBLIND\undefined
\section*{Acknowledgments}

The Students of the Graphs and Networks Computer Lab 2018/19 are, in
alphabetical order:
Yasmin Aydin,
Tarcyane Barata Garcia,
Viola Braunm{\"u}ller,
Dennis Carrer,
Sascha Phillip Clausen,
Trung Duong Duong,
Jia Fu,
Nora Grieb,
Amir Behzad Hadji Adineh,
Tobias Hagemann,
Sofia Haller,
Stephan K{\"u}hnel,
D{\'e}sir{\'e}e Langer,
Iryna Manuilova,
Paul Michaelis,
Thilo M{\"u}hl-Benninghaus,
Kolja Nenoff,
Timo Pauw,
Jens-Tilman Rau,
David Schaller,
Moritz St{\"o}cklin,
Sarah Strobel,
Christian Stur,
Justus T{\"a}ger,
Sarah Freiin von L{\"o}hneysen,
J{\"o}rg Walter.

This work was supported in part by the German Federal Ministry of Education
and Research within the project Competence Center for Scalable Data
Services and Solutions (ScaDS) Dresden/Leipzig (BMBF 01IS14014B), and the
German Research Foundation (DFG; grants STA 850/15-2).
\else
\fi

\bigskip

\par\noindent\emph{The authors have nothing to disclose.}
\goodbreak
\vspace*{1cm}
\bibliography{blubber}
\end{document}


\thispagestyle{empty} 
\begin{center}
  \large \textbf{Superbubbles as an Empirical Characteristic of Directed
    Networks}
  
  \vspace*{1cm}
  SUPPLEMENTAL MATERIAL
  
  \vspace*{2cm}
\ifx\DOUBLEBLIND\undefined
  Fabian G{\"a}rtner, Felix K{\"u}hnl, Carsten R.\ Seemann,\par 
  The Students of the Graphs and Networks Computer Lab 2018/19,\par
  Christian H{\"o}ner zu Siederdissen and Peter F.\ Stadler
\else
\fi
\end{center}
\clearpage
  
\section*{Description}

This document includes all the data analyzed in the main paper. It consists of
twelve tables containing all computed graph descriptors and metrics for four
different network datasets: the supergenome dataset (Tabs.\ 1--3), the
Stanford dataset (Tabs.\ 4--6), the LDBC dataset (Tabs.\ 7--9), and the
standard random dataset (Tabs.\ 10--12). The individual datasets are described
in the paper. Each of the three tables per dataset contains the
\emph{un}directed graph descriptors, the directed graph descriptors, and the
superbubble descriptors, respectively.

\begin{sidewaystable}
\caption{Values of the \emph{undirected} graph descriptors for the supergenome dataset.}
\begin{tabular}{|l|r|r|r|r|r|r|r|r|}
\hline
\thead{Dataset} & \thead{N} & \thead{M} & \thead{ME} & \thead{deg} & \thead{GD} & \thead{CC} & \thead{R} & \thead{SS} \\\hline
\texttt{canfam1\_3}  & 5028130 & 8959919 & 0.244154 & 6 & 3.54398e-07 & 1 & 0.414782 & 0.950578 \\
\texttt{canfam2\_4}  & 5072165 & 10021044 & 0.318041 & 8 & 3.89517e-07 & 1 & 0.342064 & 0.943163 \\
\texttt{felcat3\_4}  & 4545633 & 8025456 & 0.303981 & 8 & 3.88402e-07 & 1 & 0.123099 & 0.917051 \\
\texttt{geofor1\_7}  & 3783877 & 8343997 & 0.45908 & 14 & 5.82774e-07 & 1 & 0.252689 & 0.956434 \\
\texttt{hg38\_20}    & 26186512 & 68401246 & 0.82203 & 26 & 9.9749e-08 & 1 & 0.415601 & 0.961235 \\
\texttt{mm5\_5}      & 8567841 & 16623066 & 0.328366 & 10 & 2.26448e-07 & 1 & 0.319935 & 0.944865 \\
\texttt{mm9\_30}     & 16728118 & 45567276 & 0.732204 & 37 & 1.62839e-07 & 1 & 0.443042 & 0.907481 \\
\texttt{tarsyr2\_20} & 30368906 & 78094418 & 0.774045 & 31 & 8.46763e-08 & 61 & 0.301218 & 0.922542 \\
\hline
\end{tabular}
\end{sidewaystable}

\begin{sidewaystable}
\caption{Values of the \emph{directed} graph descriptors for the supergenome dataset.}
\begin{tabular}{|l|r|r|r|r|r|r|r|r|r|}
\hline
\thead{Dataset} & \thead{$\bm{deg_\leftarrow}$} & \thead{$\bm{deg_\rightarrow}$} & \thead{SCC} & \thead{BE} & \thead{$\bm{R_{\leftarrow\leftarrow}}$} & \thead{$\bm{R_{\leftarrow\rightarrow}}$} & \thead{$\bm{R_{\rightarrow\leftarrow}}$} & \thead{$\bm{R_{\rightarrow\rightarrow}}$} & \thead{H} \\\hline
\texttt{canfam1\_3}  & 3 & 3 & 10226 & 0.425673 & 0.285049 & 0.322407 & 0.414782 & 0.273798 & 0.0369015 \\
\texttt{canfam2\_4}  & 4 & 4 & 10825 & 0.486657 & 0.222815 & 0.277462 & 0.342064 & 0.216514 & 0.0455087 \\
\texttt{felcat3\_4}  & 4 & 4 & 42471 & 0.315417 & 0.0913772 & 0.265761 & 0.123099 & 0.0911115 & 0.0840392 \\
\texttt{geofor1\_7}  & 7 & 7 & 5328 & 0.617838 & 0.129412 & 0.218323 & 0.252689 & 0.12964 & 0.0473346 \\
\texttt{hg38\_20}    & 15 & 17 & 7268 & 0.752629 & 0.290293 & 0.350248 & 0.415601 & 0.292341 & 0.0293036 \\
\texttt{mm5\_5}      & 5 & 5 & 95229 & 0.374326 & 0.245333 & 0.341557 & 0.319935 & 0.228311 & 0.0544757 \\
\texttt{mm9\_30}     & 21 & 20 & 44122 & 0.60089 & 0.414032 & 0.45828 & 0.443042 & 0.40942 & 0.0751565 \\
\texttt{tarsyr2\_20} & 17 & 16 & 154117 & 0.657013 & 0.233599 & 0.303633 & 0.301218 & 0.233787 & 0.0568008 \\
\hline
\end{tabular}
\end{sidewaystable}

\begin{sidewaystable}
\caption{Values of the \emph{superbubble} descriptors for the supergenome dataset.}
\begin{tabular}{|l|r|r|r|r|r|r|r|r|r|r|r|r|r|r|}
\hline
\thead{Dataset} & \thead{S} & \thead{VS} & \thead{ES} & \thead{MS} & \thead{mVS} & \thead{mES} & \thead{C} & \thead{CS} & \thead{depth} & \thead{P} & \thead{PL} & \thead{aP} & \thead{aPL} & \thead{SD}  \\\hline
\texttt{canfam1\_3} & 1328348 & 0.415181 & 0.485531 & 878184 & 61 & 62 & 1004658 & 58 & 2 & 34 & 60 & 2.28371 & 2.97535 & 0.817645 \\
\texttt{canfam2\_4} & 900570 & 0.311271 & 0.346538 & 600039 & 49 & 65 & 726247 & 35 & 3 & 36869 & 48 & 3.16791 & 3.22902 & 0.817748 \\
\texttt{felcat3\_4} & 864604 & 0.405645 & 0.509091 & 348187 & 340 & 505 & 678965 & 136 & 5 & $\infty$ & 318 & $\infty$ & 3.11772 & 0.860604 \\
\texttt{geofor1\_7} & 248778 & 0.16144 & 0.213755 & 60969 & 326 & 487 & 171957 & 150 & 6 & $\infty$ & 325 & $\infty$ & 3.64778 & 0.81775 \\
\texttt{hg38\_20} & 62939 & 0.00596719 & 0.00611032 & 25985 & 109 & 162 & 56492 & 33 & 6 & 291594320 & 108 & 8203.51 & 2.73751 & 0.909297 \\
\texttt{mm5\_5} & 1202215 & 0.301006 & 0.353767 & 621905 & 176 & 264 & 911412 & 72 & 4 & $\infty$ & 175 & $\infty$ & 3.56192 & 0.783625 \\
\texttt{mm9\_30} & 349966 & 0.0500894 & 0.0501826 & 125138 & 50 & 80 & 340786 & 18 & 4 & 381363 & 49 & 6.7438 & 2.16192 & 0.976006 \\
\texttt{tarsyr2\_20} & 217149 & 0.0185623 & 0.0190412 & 58980 & 143 & 219 & 209543 & 40 & 5 & 153994257409 & 142 & 1.18295e+06 & 2.26843 & 0.964629 \\
\hline
\end{tabular}
\end{sidewaystable}

\begin{sidewaystable}
\vspace{\vspacestandford}
\caption{Values of the \emph{undirected} graph descriptors for the Stanford dataset.}
\begin{tabular}{|l|r|r|r|r|r|r|r|r|}
\hline
\thead{Dataset} & \thead{N} & \thead{M} & \thead{ME} & \thead{deg} & \thead{GD} & \thead{CC} & \thead{R} & \thead{SS} \\\hline
\texttt{CollegeMsg} & 1899 & 20296 & 0.660801 & 339 & 0.00563105 & 4 & -0.137458 & 0.671347 \\
\texttt{amazon0302} & 262111 & 1234877 & 0 & 425 & 1.79744e-05 & 1 & 0.00267724 & 0.454878 \\
\texttt{amazon0312} & 400727 & 3200440 & 0 & 2757 & 1.99303e-05 & 1 & -0.0445745 & 0.261221 \\
\texttt{amazon0505} & 410236 & 3356824 & 0 & 2770 & 1.99463e-05 & 1 & -0.0435129 & 0.276952 \\
\texttt{amazon0601} & 403394 & 3387388 & 0 & 2761 & 2.08165e-05 & 7 & -0.0435352 & 0.28163 \\
\texttt{cit-HepPh} & 34546 & 421578 & 0 & 846 & 0.00035326 & 61 & -0.00262823 & 0.498742 \\
\texttt{cit-Patents} & 3774768 & 16518948 & 0 & 793 & 1.15932e-06 & 3627 & 0.133174 & 0.501186 \\
\texttt{email-Eu-core-temporal} & 986 & 24929 & 0.924988 & 544 & 0.025668 & 1 & -0.0137082 & 0.896762 \\
\texttt{email-EuAll} & 265214 & 420045 & 0 & 7636 & 5.97179e-06 & 15836 & -0.210412 & 0.177435 \\
\texttt{gplus\_combined} & 107614 & 13673453 & 0.551615 & 22022 & 0.00118071 & 1 & 0.689601 & 0.677025 \\
\texttt{p2p-Gnutella04} & 10876 & 39994 & 0 & 103 & 0.00033814 & 1 & -0.00829953 & 0.668368 \\
\texttt{p2p-Gnutella05} & 8846 & 31839 & 0 & 88 & 0.000406925 & 3 & -0.00535846 & 0.563469 \\
\texttt{p2p-Gnutella06} & 8717 & 31525 & 0 & 115 & 0.000414926 & 1 & -0.00317719 & 0.614129 \\
\texttt{p2p-Gnutella08} & 6301 & 20777 & 0 & 97 & 0.000523399 & 2 & -0.028534 & 0.491797 \\
\texttt{p2p-Gnutella09} & 8114 & 26013 & 0 & 102 & 0.000395161 & 6 & -0.0326591 & 0.496475 \\
\texttt{p2p-Gnutella24} & 26518 & 65369 & 0 & 355 & 9.29623e-05 & 11 & -0.00558292 & 0.660619 \\
\texttt{p2p-Gnutella25} & 22687 & 54705 & 0 & 66 & 0.00010629 & 13 & -0.0062193 & 0.687883 \\
\texttt{p2p-Gnutella30} & 36682 & 88328 & 0 & 55 & 6.56454e-05 & 12 & -0.0214005 & 0.683563 \\
\texttt{p2p-Gnutella31} & 62586 & 147892 & 0 & 95 & 3.7757e-05 & 12 & -0.00628506 & 0.653816 \\
\texttt{soc-Epinions1} & 75879 & 508837 & 0 & 3079 & 8.83774e-05 & 2 & -0.0412863 & 0.574183 \\
\texttt{soc-LiveJournal1} & 4847571 & 68993773 & 0 & 22889 & 2.93604e-06 & 1876 & 0.0642377 & 0.40907 \\
\texttt{soc-Slashdot0811} & 77360 & 905468 & 0 & 5048 & 0.000151302 & 1 & -0.0488503 & 0.476566 \\
\texttt{soc-Slashdot0902} & 82168 & 948464 & 0 & 5064 & 0.000140482 & 1 & -0.0516366 & 0.470255 \\
\texttt{soc-pokec-relationships} & 1632803 & 30622564 & 0 & 20518 & 1.14861e-05 & 1 & -0.000492779 & 0.477327 \\
\texttt{soc-sign-bitcoinalpha} & 3783 & 24186 & 0 & 888 & 0.00169047 & 5 & -0.163484 & 0.659601 \\
\texttt{soc-sign-bitcoinotc} & 5881 & 35592 & 0 & 1298 & 0.00102926 & 4 & -0.163954 & 0.631987 \\
\texttt{soc-sign-epinions} & 131828 & 841372 & 0 & 3622 & 4.84146e-05 & 5816 & -0.0643017 & 0.544485 \\
\texttt{sx-askubuntu} & 159316 & 596933 & 0.381055 & 6486 & 2.35185e-05 & 4250 & -0.102788 & 0.505169 \\
\texttt{sx-mathoverflow} & 24818 & 239978 & 0.52625 & 2760 & 0.000389633 & 104 & -0.135359 & 0.727292 \\
\texttt{sx-superuser} & 194085 & 924886 & 0.359204 & 14504 & 2.45531e-05 & 3197 & -0.0813261 & 0.532765 \\
\texttt{twitter\_combined} & 81306 & 1768149 & 0.269591 & 3758 & 0.000267472 & 1 & -0.0236148 & 0.481903 \\
\texttt{web-Google} & 875713 & 5105039 & 0 & 6353 & 6.65696e-06 & 2746 & -0.0652002 & 0.0814377 \\
\texttt{web-NotreDame} & 325729 & 1497134 & 0 & 10721 & 1.41107e-05 & 1 & -0.0616567 & 0.324627 \\
\texttt{web-Stanford} & 281903 & 2312497 & 0 & 38626 & 2.90994e-05 & 365 & -0.122013 & 0.230393 \\
\texttt{wiki-Talk} & 2394385 & 5021410 & 0 & 100032 & 8.75866e-07 & 2555 & -0.0852645 & 0.228997 \\
\texttt{wiki-Vote} & 7115 & 103689 & 0 & 1167 & 0.00204854 & 24 & -0.0832446 & 0.687712 \\
\texttt{wiki-talk-temporal} & 1140149 & 3309592 & 0.577488 & 142148 & 2.54596e-06 & 47207 & -0.11657 & 0.328096 \\
\hline
\end{tabular}
\end{sidewaystable}

\begin{sidewaystable}
\vspace{\vspacestandford}
\caption{Values of the \emph{directed} graph descriptors for the Stanford dataset.}
\begin{tabular}{|l|r|r|r|r|r|r|r|r|r|}
\hline
\thead{Dataset} & \thead{$\bm{deg_\leftarrow}$} & \thead{$\bm{deg_\rightarrow}$} & \thead{SCC} & \thead{BE} & \thead{$\bm{R_{\leftarrow\leftarrow}}$} & \thead{$\bm{R_{\leftarrow\rightarrow}}$} & \thead{$\bm{R_{\rightarrow\leftarrow}}$} & \thead{$\bm{R_{\rightarrow\rightarrow}}$} & \thead{H} \\\hline
\texttt{CollegeMsg} & 137 & 237 & 293 & 0.749754 & -0.0977021 & -0.0923532 & -0.137458 & -0.145423 & 0.605004 \\
\texttt{amazon0302} & 420 & 5 & 7566 & 0.542702 & 0.0771522 & 0.0278404 & 0.00267724 & 0.102709 & 0.170112 \\
\texttt{amazon0312} & 2747 & 10 & 11511 & 0.531534 & 0.0136628 & 0.0399033 & -0.0445745 & 0.272131 & 0.259444 \\
\texttt{amazon0505} & 2760 & 10 & 11904 & 0.54658 & 0.0225778 & 0.0580517 & -0.0435129 & 0.226217 & 0.255811 \\
\texttt{amazon0601} & 2751 & 10 & 1657 & 0.55735 & 0.0244154 & 0.0567056 & -0.0435352 & 0.232733 & 0.253739 \\
\texttt{cit-HepPh} & 846 & 411 & 1399 & 0.00322123 & 0.0771375 & -0.0398235 & -0.00262823 & 0.111472 & 0.438023 \\
\texttt{cit-Patents} & 779 & 770 & 0 & 0 & 0.142261 & -0.0471524 & 0.133174 & 0.104213 & 0.261908 \\
\texttt{email-Eu-core-temporal} & 211 & 333 & 123 & 0.71122 & 0.0247314 & 0.0150836 & -0.0137082 & -0.0199169 & 0.462205 \\
\texttt{email-EuAll} & 7631 & 930 & 9189 & 0.438141 & -0.162543 & -0.145922 & -0.210412 & -0.182863 & 0.871444 \\
\texttt{gplus\_combined} & 17055 & 5056 & 13849 & 0.20992 & -0.043031 & 0.0188671 & -0.0742248 & 0.0573339 & 0.819481 \\
\texttt{p2p-Gnutella04} & 72 & 100 & 2210 & 0 & 0.0042255 & -0.0116766 & -0.00829953 & -0.00366304 & 0.295528 \\
\texttt{p2p-Gnutella05} & 79 & 65 & 1714 & 0 & 0.0311929 & -0.00017749 & -0.00535846 & -0.00167642 & 0.294215 \\
\texttt{p2p-Gnutella06} & 64 & 113 & 1661 & 0 & 0.0879897 & 0.0322133 & -0.00317719 & 0.00817549 & 0.295178 \\
\texttt{p2p-Gnutella08} & 91 & 48 & 1178 & 0 & 0.107873 & 0.0315348 & -0.028534 & -0.0136803 & 0.328442 \\
\texttt{p2p-Gnutella09} & 92 & 61 & 1557 & 0 & 0.104187 & 0.0189707 & -0.0326591 & -0.00623989 & 0.331892 \\
\texttt{p2p-Gnutella24} & 355 & 79 & 4423 & 0 & 0.0116544 & -0.00320436 & -0.00558292 & 0.004415 & 0.327877 \\
\texttt{p2p-Gnutella25} & 36 & 64 & 3624 & 0 & -0.0058195 & -0.0024812 & -0.0062193 & 0.000112338 & 0.325374 \\
\texttt{p2p-Gnutella30} & 54 & 54 & 5699 & 0 & 0.0440849 & 0.0237749 & -0.0214005 & -0.00855574 & 0.341343 \\
\texttt{p2p-Gnutella31} & 68 & 78 & 9581 & 0 & 0.0345039 & 0.00756586 & -0.00628506 & -0.00305984 & 0.349917 \\
\texttt{soc-Epinions1} & 3035 & 1801 & 8431 & 0.0176756 & 0.0424935 & 0.073024 & -0.0412863 & -0.0101631 & 0.638096 \\
\texttt{soc-LiveJournal1} & 13906 & 20293 & 461640 & 0.0276269 & 0.0636388 & 0.12009 & 0.0213629 & 0.0477492 & 0.544738 \\
\texttt{soc-Slashdot0811} & 2540 & 2508 & 3595 & 0.878296 & -0.0476222 & -0.0426242 & -0.0488503 & -0.0456693 & 0.554386 \\
\texttt{soc-Slashdot0902} & 2553 & 2511 & 6928 & 0.854186 & -0.0495735 & -0.0413662 & -0.0516366 & -0.0450336 & 0.548488 \\
\texttt{soc-pokec-relationships} & 13733 & 8763 & 154972 & 0.543429 & 0.000586022 & 0.00252156 & -0.000492779 & -0.000425007 & 0.468531 \\
\texttt{soc-sign-bitcoinalpha} & 398 & 490 & 342 & 1.41251 & -0.161702 & -0.154611 & -0.163484 & -0.15639 & 0.880293 \\
\texttt{soc-sign-bitcoinotc} & 535 & 763 & 602 & 0.792313 & -0.161213 & -0.148855 & -0.163954 & -0.151106 & 0.91547 \\
\texttt{soc-sign-epinions} & 3478 & 2070 & 13525 & 0.485062 & 0.00512147 & 0.0444978 & -0.0643017 & -0.0305167 & 0.606906 \\
\texttt{sx-askubuntu} & 1954 & 4966 & 25067 & 0.0953474 & -0.0886782 & -0.0867718 & -0.102788 & -0.0956014 & 0.681661 \\
\texttt{sx-mathoverflow} & 969 & 1849 & 4671 & 0.383356 & -0.106931 & -0.103006 & -0.135359 & -0.127363 & 0.859381 \\
\texttt{sx-superuser} & 2513 & 14255 & 28502 & 0.378558 & -0.0743548 & -0.0694729 & -0.0813261 & -0.0683881 & 0.724175 \\
\texttt{twitter\_combined} & 3383 & 1205 & 7601 & 0.481686 & 0.000198925 & 0.0550662 & -0.0236148 & 0.0641385 & 0.612801 \\
\texttt{web-Google} & 6326 & 456 & 60793 & 0.306751 & -0.0137601 & 0.0331156 & -0.0652002 & 0.0583786 & 0.506924 \\
\texttt{web-NotreDame} & 10721 & 3445 & 20015 & 0.428944 & -0.0227764 & 0.256693 & -0.0616567 & -0.0135109 & 0.660539 \\
\texttt{web-Stanford} & 38606 & 255 & 5223 & 0.276637 & -0.0119076 & 0.00854007 & -0.122013 & 0.045759 & 0.677823 \\
\texttt{wiki-Talk} & 3311 & 100022 & 53627 & 0.0233279 & -0.0565771 & -0.0481652 & -0.0852645 & -0.0572358 & 0.929567 \\
\texttt{wiki-Vote} & 457 & 893 & 421 & 0.0564573 & 0.00509101 & 0.0070958 & -0.0832446 & -0.0189092 & 0.543591 \\
\texttt{wiki-talk-temporal} & 3316 & 141884 & 74022 & 0.0341386 & -0.0466978 & -0.0245695 & -0.11657 & -0.0407581 & 0.882692 \\
\hline
\end{tabular}
\end{sidewaystable}

\begin{sidewaystable}
\vspace{\vspacestandford}
\caption{Values of the \emph{superbubble} descriptors for the Stanford dataset.}
\begin{tabular}{|l|r|r|r|r|r|r|r|r|r|r|r|r|r|r|}
\hline
\thead{Dataset} & \thead{S} & \thead{VS} & \thead{ES} & \thead{MS} & \thead{mVS} & \thead{mES} & \thead{C} & \thead{CS} & \thead{depth} & \thead{P} & \thead{PL} & \thead{aP} & \thead{aPL} & \thead{SD}  \\\hline
\texttt{CollegeMsg} & 3 & 0.00315956 & 0.00157978 & 3 & 2 & 1 & 3 & 1 & 1 & 1 & 1 & 0 & 0 & 0 \\
\texttt{amazon0302} & 13 & 9.91946e-05 & 4.95973e-05 & 13 & 2 & 1 & 13 & 1 & 1 & 1 & 1 & 0 & 0 & 0 \\
\texttt{amazon0312} & 23 & 0.000114791 & 5.73957e-05 & 23 & 2 & 1 & 23 & 1 & 1 & 1 & 1 & 0 & 0 & 0 \\
\texttt{amazon0505} & 18 & 8.77544e-05 & 4.38772e-05 & 18 & 2 & 1 & 18 & 1 & 1 & 1 & 1 & 0 & 0 & 0 \\
\texttt{amazon0601} & 3 & 1.48738e-05 & 7.4369e-06 & 3 & 2 & 1 & 3 & 1 & 1 & 1 & 1 & 0 & 0 & 0 \\
\texttt{cit-HepPh} & 135 & 0.00787356 & 0.00439993 & 128 & 4 & 6 & 135 & 1 & 1 & 4 & 3 & 2.28571 & 2.14286 & 1 \\
\texttt{cit-Patents} & 9457 & 0.00499368 & 0.00253181 & 9415 & 7 & 13 & 9457 & 1 & 1 & 18 & 6 & 2.45238 & 2.09524 & 0.979819 \\
\texttt{email-Eu-core-temporal} & 1 & 0.0020284 & 0.0010142 & 1 & 2 & 1 & 1 & 1 & 1 & 1 & 1 & 0 & 0 & 0 \\
\texttt{email-EuAll} & 13285 & 0.100183 & 0.0500916 & 13285 & 2 & 1 & 13285 & 1 & 1 & 1 & 1 & 0 & 0 & 0 \\
\texttt{gplus\_combined} & 0 & 0 & 0 & 0 & 0 & 0 & 0 & 0 & 0 & 0 & 0 & 0 & 0 & 0 \\
\texttt{p2p-Gnutella04} & 100 & 0.0183891 & 0.00919456 & 100 & 2 & 1 & 100 & 1 & 1 & 1 & 1 & 0 & 0 & 0 \\
\texttt{p2p-Gnutella05} & 62 & 0.0140176 & 0.00700882 & 62 & 2 & 1 & 62 & 1 & 1 & 1 & 1 & 0 & 0 & 0 \\
\texttt{p2p-Gnutella06} & 52 & 0.011816 & 0.00596536 & 52 & 2 & 1 & 52 & 1 & 1 & 1 & 1 & 0 & 0 & 0 \\
\texttt{p2p-Gnutella08} & 44 & 0.013966 & 0.00698302 & 44 & 2 & 1 & 44 & 1 & 1 & 1 & 1 & 0 & 0 & 0 \\
\texttt{p2p-Gnutella09} & 65 & 0.0158984 & 0.00801085 & 65 & 2 & 1 & 65 & 1 & 1 & 1 & 1 & 0 & 0 & 0 \\
\texttt{p2p-Gnutella24} & 117 & 0.00874877 & 0.0044121 & 117 & 2 & 1 & 117 & 1 & 1 & 1 & 1 & 0 & 0 & 0 \\
\texttt{p2p-Gnutella25} & 85 & 0.00749328 & 0.00374664 & 85 & 2 & 1 & 85 & 1 & 1 & 1 & 1 & 0 & 0 & 0 \\
\texttt{p2p-Gnutella30} & 108 & 0.00588845 & 0.00294422 & 108 & 2 & 1 & 108 & 1 & 1 & 1 & 1 & 0 & 0 & 0 \\
\texttt{p2p-Gnutella31} & 260 & 0.00829259 & 0.00415428 & 260 & 2 & 1 & 260 & 1 & 1 & 1 & 1 & 0 & 0 & 0 \\
\texttt{soc-Epinions1} & 824 & 0.021587 & 0.0108858 & 823 & 3 & 3 & 824 & 1 & 1 & 2 & 2 & 2 & 2 & 1 \\
\texttt{soc-LiveJournal1} & 5432 & 0.00223225 & 0.001135 & 5399 & 4 & 5 & 5432 & 1 & 1 & 3 & 3 & 2.06061 & 2.0303 & 0.989899 \\
\texttt{soc-Slashdot0811} & 0 & 0 & 0 & 0 & 0 & 0 & 0 & 0 & 0 & 0 & 0 & 0 & 0 & 0 \\
\texttt{soc-Slashdot0902} & 0 & 0 & 0 & 0 & 0 & 0 & 0 & 0 & 0 & 0 & 0 & 0 & 0 & 0 \\
\texttt{soc-pokec-relationships} & 1468 & 0.00179385 & 0.000901517 & 1466 & 3 & 3 & 1468 & 1 & 1 & 2 & 2 & 2 & 2 & 1 \\
\texttt{soc-sign-bitcoinalpha} & 5 & 0.0026434 & 0.0013217 & 5 & 2 & 1 & 5 & 1 & 1 & 1 & 1 & 0 & 0 & 0 \\
\texttt{soc-sign-bitcoinotc} & 4 & 0.00136031 & 0.000680156 & 4 & 2 & 1 & 4 & 1 & 1 & 1 & 1 & 0 & 0 & 0 \\
\texttt{soc-sign-epinions} & 5961 & 0.0897078 & 0.0453546 & 5952 & 3 & 3 & 5961 & 1 & 1 & 2 & 2 & 2 & 2 & 1 \\
\texttt{sx-askubuntu} & 2168 & 0.0271159 & 0.0136207 & 2167 & 3 & 3 & 2168 & 1 & 1 & 2 & 2 & 2 & 2 & 1 \\
\texttt{sx-mathoverflow} & 41 & 0.00330405 & 0.00165203 & 41 & 2 & 1 & 41 & 1 & 1 & 1 & 1 & 0 & 0 & 0 \\
\texttt{sx-superuser} & 1668 & 0.0171523 & 0.00859417 & 1668 & 2 & 1 & 1668 & 1 & 1 & 1 & 1 & 0 & 0 & 0 \\
\texttt{twitter\_combined} & 0 & 0 & 0 & 0 & 0 & 0 & 0 & 0 & 0 & 0 & 0 & 0 & 0 & 0 \\
\texttt{web-Google} & 6477 & 0.015472 & 0.00923819 & 5766 & 15 & 27 & 6477 & 1 & 1 & 16 & 9 & 2.13783 & 2.05063 & 0.985761 \\
\texttt{web-NotreDame} & 1714 & 0.00959693 & 0.00580544 & 1691 & 16 & 120 & 1713 & 2 & 2 & 16384 & 15 & 714.652 & 2.73913 & 0.957971 \\
\texttt{web-Stanford} & 933 & 0.00615105 & 0.00345864 & 915 & 7 & 9 & 933 & 1 & 1 & 4 & 5 & 2.11111 & 2.16667 & 0.968254 \\
\texttt{wiki-Talk} & 4737 & 0.00395049 & 0.00197921 & 4736 & 3 & 3 & 4737 & 1 & 1 & 2 & 2 & 2 & 2 & 1 \\
\texttt{wiki-Vote} & 23 & 0.00646521 & 0.00323261 & 23 & 2 & 1 & 23 & 1 & 1 & 1 & 1 & 0 & 0 & 0 \\
\texttt{wiki-talk-temporal} & 1253 & 0.00219533 & 0.00109898 & 1253 & 2 & 1 & 1253 & 1 & 1 & 1 & 1 & 0 & 0 & 0 \\
\hline
\end{tabular}
\end{sidewaystable}

\begin{sidewaystable}
\vspace{\vspaceldbc}
\caption{Values of the \emph{undirected} graph descriptors for the LDBC dataset.}
\begin{tabular}{|l|r|r|r|r|r|r|r|r|}
\hline
\thead{Dataset} & \thead{N} & \thead{M} & \thead{ME} & \thead{deg} & \thead{GD} & \thead{CC} & \thead{R} & \thead{SS} \\\hline
\texttt{datagen-7\_5-fb} & 633432 & 34185747 & 0 & 2772 & 8.52012e-05 & 1 & 0.194012 & 0.245852 \\
\texttt{datagen-7\_6-fb} & 754147 & 42162988 & 0 & 2899 & 7.41344e-05 & 1 & 0.206316 & 0.248139 \\
\texttt{datagen-7\_7-zf} & 13180508 & 32791267 & 0 & 2313 & 1.88753e-07 & 199019 & -0.1256 & 0.0816684 \\
\texttt{datagen-7\_8-zf} & 16521886 & 41025255 & 0 & 2354 & 1.50291e-07 & 253053 & -0.119763 & 0.0806329 \\
\texttt{datagen-7\_9-fb} & 1387587 & 85670523 & 0 & 3201 & 4.4495e-05 & 1 & 0.263505 & 0.251526 \\
\texttt{datagen-8\_0-fb} & 1706561 & 107507376 & 0 & 3310 & 3.69143e-05 & 1 & 0.276152 & 0.255602 \\
\texttt{datagen-8\_1-fb} & 2072117 & 134267822 & 0 & 3408 & 3.12711e-05 & 1 & 0.291181 & 0.257618 \\
\texttt{datagen-8\_2-zf} & 43734497 & 106440188 & 0 & 2283 & 5.5649e-08 & 669856 & 0.038178 & 0.0839126 \\
\texttt{datagen-8\_3-zf} & 53525014 & 130579909 & 0 & 2101 & 4.55788e-08 & 869190 & 0.0981702 & 0.0821893 \\
\texttt{datagen-8\_4-fb} & 3809084 & 269479177 & 0 & 3750 & 1.85731e-05 & 1 & 0.324805 & 0.269613 \\
\texttt{dota-league} & 61170 & 50870313 & 0 & 17004 & 0.0135955 & 1 & 1.90665 & 0.782976 \\
\texttt{graph500-22} & 2396657 & 64155735 & 0 & 162768 & 1.11692e-05 & 734 & 0.0791385 & 0.543183 \\
\texttt{graph500-23} & 4610222 & 129333677 & 0 & 256708 & 6.0851e-06 & 1548 & 0.0680088 & 0.527994 \\
\texttt{graph500-24} & 8870942 & 260379520 & 0 & 406416 & 3.30878e-06 & 2901 & 0.0582805 & 0.511917 \\
\hline
\end{tabular}
\end{sidewaystable}

\begin{sidewaystable}
\vspace{\vspaceldbc}
\caption{Values of the \emph{directed} graph descriptors for the LDBC dataset.}
\begin{tabular}{|l|r|r|r|r|r|r|r|r|r|}
\hline
\thead{Dataset} & \thead{$\bm{deg_\leftarrow}$} & \thead{$\bm{deg_\rightarrow}$} & \thead{SCC} & \thead{BE} & \thead{$\bm{R_{\leftarrow\leftarrow}}$} & \thead{$\bm{R_{\leftarrow\rightarrow}}$} & \thead{$\bm{R_{\rightarrow\leftarrow}}$} & \thead{$\bm{R_{\rightarrow\rightarrow}}$} & \thead{H} \\\hline
\texttt{datagen-7\_5-fb} & 2597 & 2706 & 0 & 0 & -0.106357 & -0.148265 & -0.184433 & -0.119919 & 0.820011 \\
\texttt{datagen-7\_6-fb} & 2752 & 2850 & 0 & 0 & -0.104827 & -0.146698 & -0.182034 & -0.117996 & 0.821265 \\
\texttt{datagen-7\_7-zf} & 1769 & 1941 & 0 & 0 & -0.0708821 & -0.065293 & -0.1256 & -0.103621 & 0.692345 \\
\texttt{datagen-7\_8-zf} & 1388 & 1885 & 0 & 0 & -0.0684085 & -0.0611063 & -0.119763 & -0.0967497 & 0.689245 \\
\texttt{datagen-7\_9-fb} & 3023 & 3089 & 0 & 0 & -0.102425 & -0.14036 & -0.177541 & -0.117966 & 0.80739 \\
\texttt{datagen-8\_0-fb} & 3310 & 3241 & 0 & 0 & -0.101964 & -0.137096 & -0.176413 & -0.117985 & 0.805318 \\
\texttt{datagen-8\_1-fb} & 3274 & 3232 & 0 & 0 & -0.10209 & -0.134706 & -0.175564 & -0.11954 & 0.801927 \\
\texttt{datagen-8\_2-zf} & 1882 & 2252 & 0 & 0 & -0.0574927 & -0.0507436 & -0.098467 & -0.0819972 & 0.682754 \\
\texttt{datagen-8\_3-zf} & 1391 & 1937 & 0 & 0 & -0.055267 & -0.048098 & -0.0943121 & -0.0781368 & 0.674608 \\
\texttt{datagen-8\_4-fb} & 3593 & 3477 & 0 & 0 & -0.0973633 & -0.124215 & -0.170645 & -0.120632 & 0.778151 \\
\texttt{dota-league} & 10038 & 13584 & 0 & 0 & 0.0736484 & -0.238497 & -0.035109 & 0.09171 & 0.817055 \\
\texttt{graph500-22} & 50447 & 162768 & 0 & 0 & -0.0391452 & -0.0552265 & -0.0670958 & -0.0242196 & 1.05559 \\
\texttt{graph500-23} & 78915 & 256708 & 0 & 0 & -0.0360313 & -0.0494488 & -0.0597288 & -0.0219937 & 1.07196 \\
\texttt{graph500-24} & 123270 & 406416 & 0 & 0 & -0.0329966 & -0.0442583 & -0.0530081 & -0.0198466 & 1.08716 \\
\hline
\end{tabular}
\end{sidewaystable}

\begin{sidewaystable}
\vspace{\vspaceldbc}
\caption{Values of the \emph{superbubble} descriptors for the LDBC dataset.}
\begin{tabular}{|l|r|r|r|r|r|r|r|r|r|r|r|r|r|r|}
\hline
\thead{Dataset} & \thead{S} & \thead{VS} & \thead{ES} & \thead{MS} & \thead{mVS} & \thead{mES} & \thead{C} & \thead{CS} & \thead{depth} & \thead{P} & \thead{PL} & \thead{aP} & \thead{aPL} & \thead{SD}  \\\hline
\texttt{datagen-7\_5-fb} & 0 & 0 & 0 & 0 & 0 & 0 & 0 & 0 & 0 & 0 & 0 & 0 & 0 & 0 \\
\texttt{datagen-7\_6-fb} & 0 & 0 & 0 & 0 & 0 & 0 & 0 & 0 & 0 & 0 & 0 & 0 & 0 & 0 \\
\texttt{datagen-7\_7-zf} & 393239 & 0.0571091 & 0.0298964 & 392841 & 4 & 5 & 393233 & 2 & 2 & 3 & 3 & 2.00503 & 2.0201 & 0.984925 \\
\texttt{datagen-7\_8-zf} & 487448 & 0.0566572 & 0.0295588 & 487002 & 5 & 5 & 487434 & 3 & 2 & 3 & 4 & 2.00224 & 2.03363 & 0.971973 \\
\texttt{datagen-7\_9-fb} & 0 & 0 & 0 & 0 & 0 & 0 & 0 & 0 & 0 & 0 & 0 & 0 & 0 & 0 \\
\texttt{datagen-8\_0-fb} & 1 & 1.17195e-06 & 5.85974e-07 & 1 & 2 & 1 & 1 & 1 & 1 & 1 & 1 & 0 & 0 & 0 \\
\texttt{datagen-8\_1-fb} & 4 & 3.86079e-06 & 1.93039e-06 & 4 & 2 & 1 & 4 & 1 & 1 & 1 & 1 & 0 & 0 & 0 \\
\texttt{datagen-8\_2-zf} & 1274414 & 0.0558878 & 0.0291964 & 1273207 & 5 & 5 & 1274385 & 3 & 2 & 3 & 4 & 2.00331 & 2.02568 & 0.977216 \\
\texttt{datagen-8\_3-zf} & 1691574 & 0.0603148 & 0.031671 & 1689810 & 6 & 6 & 1691528 & 4 & 2 & 3 & 5 & 2.0017 & 2.02664 & 0.976323 \\
\texttt{datagen-8\_4-fb} & 6 & 3.15036e-06 & 1.57518e-06 & 6 & 2 & 1 & 6 & 1 & 1 & 1 & 1 & 0 & 0 & 0 \\
\texttt{dota-league} & 0 & 0 & 0 & 0 & 0 & 0 & 0 & 0 & 0 & 0 & 0 & 0 & 0 & 0 \\
\texttt{graph500-22} & 4473 & 0.00373145 & 0.00186635 & 4473 & 2 & 1 & 4473 & 1 & 1 & 1 & 1 & 0 & 0 & 0 \\
\texttt{graph500-23} & 8219 & 0.00356425 & 0.00178278 & 8219 & 2 & 1 & 8219 & 1 & 1 & 1 & 1 & 0 & 0 & 0 \\
\texttt{graph500-24} & 15288 & 0.00344597 & 0.00172338 & 15288 & 2 & 1 & 15288 & 1 & 1 & 1 & 1 & 0 & 0 & 0 \\
\hline
\end{tabular}
\end{sidewaystable}

\begin{sidewaystable}
\vspace{\vspacestdrand}
\caption{Values of the \emph{undirected} graph descriptors for the standard random dataset.}
\begin{tabular}{|l|r|r|r|r|r|r|r|r|}
\hline
\thead{Dataset} & \thead{N} & \thead{M} & \thead{ME} & \thead{deg} & \thead{GD} & \thead{CC} & \thead{R} & \thead{SS} \\\hline
\texttt{BA1} & 100000 & 99999 & 0 & 469 & 1e-05 & 1 & -nan & 0.125639 \\
\texttt{BA2} & 100000 & 99999 & 0 & 523 & 1e-05 & 1 & -nan & 0.134829 \\
\texttt{BA3} & 100000 & 99999 & 0 & 736 & 1e-05 & 1 & -nan & 0.152272 \\
\texttt{ER\_0.01\_1} & 100000 & 100002684 & 0 & 2222 & 0.0100004 & 1 & 1014.73 & 0.999582 \\
\texttt{ER\_0.01\_2} & 100000 & 99993010 & 0 & 2184 & 0.0099994 & 1 & 1012 & 0.999565 \\
\texttt{ER\_0.01\_3} & 100000 & 99997576 & 0 & 2198 & 0.00999986 & 1 & 1011.55 & 0.999564 \\
\texttt{ER\_0.05\_1} & 100000 & 500017674 & 0 & 10442 & 0.0500023 & 1 & 5265.41 & 1.00013 \\
\texttt{ER\_0.05\_2} & 100000 & 499995047 & 0 & 10426 & 0.05 & 1 & 5258.51 & 1.00011 \\
\texttt{ER\_0.05\_3} & 100000 & 500012912 & 0 & 10398 & 0.0500018 & 1 & 5267.39 & 1.00012 \\
\texttt{ER\_0.10\_1} & 100000 & 1000000234 & 0 & 20540 & 0.100001 & 1 & 11162.4 & 1.00029 \\
\texttt{ER\_0.10\_2} & 100000 & 999987214 & 0 & 20591 & 0.0999997 & 1 & 11124.9 & 1.00028 \\
\texttt{ER\_0.10\_3} & 100000 & 999946928 & 0 & 20544 & 0.0999957 & 1 & 11163 & 1.00028 \\
\texttt{WS\_0.01\_1} & 100000 & 200000 & 0 & 6 & 2.00002e-05 & 1 & -nan & 0.997499 \\
\texttt{WS\_0.01\_2} & 100000 & 200000 & 0 & 6 & 2.00002e-05 & 1 & -nan & 0.99747 \\
\texttt{WS\_0.01\_3} & 100000 & 200000 & 0 & 6 & 2.00002e-05 & 1 & -nan & 0.997488 \\
\texttt{WS\_0.05\_1} & 100000 & 200000 & 0 & 7 & 2.00002e-05 & 1 & -nan & 0.987955 \\
\texttt{WS\_0.05\_2} & 100000 & 200000 & 0 & 7 & 2.00002e-05 & 1 & -nan & 0.987782 \\
\texttt{WS\_0.05\_3} & 100000 & 200000 & 0 & 9 & 2.00002e-05 & 1 & -nan & 0.987969 \\
\texttt{WS\_0.10\_1} & 100000 & 200000 & 0 & 8 & 2.00002e-05 & 1 & -nan & 0.977286 \\
\texttt{WS\_0.10\_2} & 100000 & 200000 & 0 & 8 & 2.00002e-05 & 1 & -nan & 0.977375 \\
\texttt{WS\_0.10\_3} & 100000 & 200000 & 0 & 8 & 2.00002e-05 & 1 & -nan & 0.977202 \\
\hline
\end{tabular}
\end{sidewaystable}

\begin{sidewaystable}
\vspace{\vspacestdrand}
\caption{Values of the \emph{directed} graph descriptors for the standard random dataset.}
\begin{tabular}{|l|r|r|r|r|r|r|r|r|r|}
\hline
\thead{Dataset} & \thead{$\bm{deg_\leftarrow}$} & \thead{$\bm{deg_\rightarrow}$} & \thead{SCC} & \thead{BE} & \thead{$\bm{R_{\leftarrow\leftarrow}}$} & \thead{$\bm{R_{\leftarrow\rightarrow}}$} & \thead{$\bm{R_{\rightarrow\leftarrow}}$} & \thead{$\bm{R_{\rightarrow\rightarrow}}$} & \thead{H} \\\hline
\texttt{BA1} & 468 & 1 & 0 & 0 & 0.141871 & -0.0342209 & -nan & -nan & 0.330609 \\
\texttt{BA2} & 523 & 1 & 0 & 0 & 0.147098 & -0.0848904 & -nan & -nan & 0.330121 \\
\texttt{BA3} & 736 & 1 & 0 & 0 & 0.136788 & -0.0872301 & -nan & -nan & 0.32713 \\
\texttt{ER\_0.01\_1} & 1135 & 1131 & 1 & 0.00998951 & 0.000122002 & -0.000110503 & -5.80301e-05 & 5.12891e-05 & 0.000495969 \\
\texttt{ER\_0.01\_2} & 1131 & 1149 & 1 & 0.00997998 & 3.49024e-05 & 0.000100475 & -5.67775e-05 & -0.000100906 & 0.000496543 \\
\texttt{ER\_0.01\_3} & 1170 & 1148 & 1 & 0.00999958 & -3.64618e-06 & 6.49632e-05 & -6.55851e-05 & 4.70956e-05 & 0.000497635 \\
\texttt{ER\_0.05\_1} & 5281 & 5312 & 1 & 0.0499833 & -7.69055e-05 & -5.50305e-05 & -9.31675e-05 & 5.22855e-05 & 9.55933e-05 \\
\texttt{ER\_0.05\_2} & 5291 & 5282 & 1 & 0.0500012 & 3.50629e-05 & -4.86547e-05 & -4.60684e-05 & -1.67444e-05 & 9.57124e-05 \\
\texttt{ER\_0.05\_3} & 5280 & 5278 & 1 & 0.0499931 & 6.53874e-05 & 8.85277e-05 & -9.8679e-05 & -3.50254e-05 & 9.55385e-05 \\
\texttt{ER\_0.10\_1} & 10370 & 10435 & 1 & 0.0999975 & -1.59101e-05 & -1.20243e-05 & 1.7856e-05 & -2.9967e-05 & 4.50789e-05 \\
\texttt{ER\_0.10\_2} & 10412 & 10422 & 1 & 0.0999753 & -2.03435e-05 & 6.91282e-06 & -3.52972e-05 & 3.79245e-05 & 4.52354e-05 \\
\texttt{ER\_0.10\_3} & 10400 & 10396 & 1 & 0.0999938 & -2.00339e-05 & -2.39655e-05 & 1.0596e-05 & -4.44829e-05 & 4.5077e-05 \\
\texttt{WS\_0.01\_1} & 4 & 2 & 515 & 0.98003 & -0.000373468 & -nan & -nan & -nan & 0.00270612 \\
\texttt{WS\_0.01\_2} & 4 & 2 & 533 & 0.97966 & 0.00208708 & -nan & -nan & -nan & 0.00272828 \\
\texttt{WS\_0.01\_3} & 4 & 2 & 435 & 0.97992 & 0.00234856 & -nan & -nan & -nan & 0.00271359 \\
\texttt{WS\_0.05\_1} & 5 & 2 & 1936 & 0.90272 & -0.00236275 & -nan & -nan & -nan & 0.0123803 \\
\texttt{WS\_0.05\_2} & 5 & 2 & 1943 & 0.90075 & -0.00405744 & -nan & -nan & -nan & 0.012573 \\
\texttt{WS\_0.05\_3} & 7 & 2 & 1976 & 0.9027 & -0.00028211 & -nan & -nan & -nan & 0.0123837 \\
\texttt{WS\_0.10\_1} & 6 & 2 & 3027 & 0.81154 & -0.00364385 & -nan & -nan & -nan & 0.0222144 \\
\texttt{WS\_0.10\_2} & 6 & 2 & 3134 & 0.81122 & -0.000499886 & -nan & -nan & -nan & 0.0221405 \\
\texttt{WS\_0.10\_3} & 6 & 2 & 3121 & 0.81051 & 0.00372439 & -nan & -nan & -nan & 0.0223861 \\
\hline
\end{tabular}
\end{sidewaystable}

\begin{sidewaystable}
\vspace{\vspacestdrand}
\caption{Values of the \emph{superbubble} descriptors for the standard random dataset.}
\begin{tabular}{|l|r|r|r|r|r|r|r|r|r|r|r|r|r|r|r|r|}
\hline
\thead{Dataset} & \thead{S} & \thead{VS} & \thead{ES} & \thead{MS} & \thead{VCS} & \thead{ECS} & \thead{mVS} & \thead{mES} & \thead{C} & \thead{CS} & \thead{depth} & \thead{P} & \thead{PL} & \thead{aP} & \thead{aPL} & \thead{SD} \\\hline
\texttt{BA1} & 16538 & 0.30892 & 0.165382 & 16538 & 0 & 0 & 2 & 1 & 16538 & 1 & 1 & 1 & 1 & 0 & 0 & 0 \\
\texttt{BA2} & 16652 & 0.31112 & 0.166522 & 16652 & 0 & 0 & 2 & 1 & 16652 & 1 & 1 & 1 & 1 & 0 & 0 & 0 \\
\texttt{BA3} & 16630 & 0.31063 & 0.166302 & 16630 & 0 & 0 & 2 & 1 & 16630 & 1 & 1 & 1 & 1 & 0 & 0 & 0 \\
\texttt{ER\_0.01\_1} & 0 & 0 & 0 & 0 & 0 & 0 & 0 & 0 & 0 & 0 & 0 & 0 & 0 & 0 & 0 & 0 \\
\texttt{ER\_0.01\_2} & 0 & 0 & 0 & 0 & 0 & 0 & 0 & 0 & 0 & 0 & 0 & 0 & 0 & 0 & 0 & 0 \\
\texttt{ER\_0.01\_3} & 0 & 0 & 0 & 0 & 0 & 0 & 0 & 0 & 0 & 0 & 0 & 0 & 0 & 0 & 0 & 0 \\
\texttt{ER\_0.05\_1} & 0 & 0 & 0 & 0 & 0 & 0 & 0 & 0 & 0 & 0 & 0 & 0 & 0 & 0 & 0 & 0 \\
\texttt{ER\_0.05\_2} & 0 & 0 & 0 & 0 & 0 & 0 & 0 & 0 & 0 & 0 & 0 & 0 & 0 & 0 & 0 & 0 \\
\texttt{ER\_0.05\_3} & 0 & 0 & 0 & 0 & 0 & 0 & 0 & 0 & 0 & 0 & 0 & 0 & 0 & 0 & 0 & 0 \\
\texttt{ER\_0.10\_1} & 0 & 0 & 0 & 0 & 0 & 0 & 0 & 0 & 0 & 0 & 0 & 0 & 0 & 0 & 0 & 0 \\
\texttt{ER\_0.10\_2} & 0 & 0 & 0 & 0 & 0 & 0 & 0 & 0 & 0 & 0 & 0 & 0 & 0 & 0 & 0 & 0 \\
\texttt{ER\_0.10\_3} & 0 & 0 & 0 & 0 & 0 & 0 & 0 & 0 & 0 & 0 & 0 & 0 & 0 & 0 & 0 & 0 \\
\texttt{WS\_0.01\_1} & 0 & 0 & 0 & 0 & 0 & 0 & 0 & 0 & 0 & 0 & 0 & 0 & 0 & 0 & 0 & 0 \\
\texttt{WS\_0.01\_2} & 0 & 0 & 0 & 0 & 0 & 0 & 0 & 0 & 0 & 0 & 0 & 0 & 0 & 0 & 0 & 0 \\
\texttt{WS\_0.01\_3} & 0 & 0 & 0 & 0 & 0 & 0 & 0 & 0 & 0 & 0 & 0 & 0 & 0 & 0 & 0 & 0 \\
\texttt{WS\_0.05\_1} & 0 & 0 & 0 & 0 & 0 & 0 & 0 & 0 & 0 & 0 & 0 & 0 & 0 & 0 & 0 & 0 \\
\texttt{WS\_0.05\_2} & 0 & 0 & 0 & 0 & 0 & 0 & 0 & 0 & 0 & 0 & 0 & 0 & 0 & 0 & 0 & 0 \\
\texttt{WS\_0.05\_3} & 0 & 0 & 0 & 0 & 0 & 0 & 0 & 0 & 0 & 0 & 0 & 0 & 0 & 0 & 0 & 0 \\
\texttt{WS\_0.10\_1} & 0 & 0 & 0 & 0 & 0 & 0 & 0 & 0 & 0 & 0 & 0 & 0 & 0 & 0 & 0 & 0 \\
\texttt{WS\_0.10\_2} & 0 & 0 & 0 & 0 & 0 & 0 & 0 & 0 & 0 & 0 & 0 & 0 & 0 & 0 & 0 & 0 \\
\texttt{WS\_0.10\_3} & 0 & 0 & 0 & 0 & 0 & 0 & 0 & 0 & 0 & 0 & 0 & 0 & 0 & 0 & 0 & 0 \\
\hline
\end{tabular}
\end{sidewaystable}